\documentclass[useAMS,usenatbib]{mn2e}
\usepackage{graphics,epsfig,psfig}
\usepackage[normalem]{ulem}
\usepackage[dvipsnames]{color}
\usepackage[]{inputenc,amssymb}
\usepackage{amsmath}
\usepackage{color}
\usepackage{textcomp}

\usepackage[toc,page]{appendix}

\def \be{\begin{equation}}
\def \ee{\end{equation}}
\def \bea{\begin{eqnarray}}
\def \eea{\end{eqnarray}}
\def\etal{{et al.}}

\def \pra{{\sf Prior-1} }
\def \prb{{\sf Prior-2} }

\def \fg{$f_{\rm gas}$ }
\def \ftemp{$f_{\rm Temp}$ }
\def \alpg{$\alpha_{\rm gas}$ }
\def \gmag{$\gamma_{\rm gas}$ }

\title[CGM Cross-correlations]{Probing the circumgalactic baryons through cross-correlations}

\author [Singh, Majumdar, Nath, Refregier and Silk]
{Priyanka Singh$^{1}$ \thanks{priyankas@rri.res.in}, 
Subhabrata Majumdar$^2$, 
 Biman B. Nath$^1$, 
 Alexandre Refregier$^3$ 
and \newauthor{Joseph Silk$^{4,5,6,7}$} 
 \\
$^1$ Raman Research Institute, Bangalore, India, 560080 \\
$^2$ Tata Institute of Fundamental Research, Mumbai, India, 400005\\
$^3$  Institute for Astronomy, Department of Physics, ETH Z\"{u}rich, Wolfgang-Pauli-Strasse 27, CH-8093 Z\"{u}rich, Switzerland\\
$^{4}$ Institut d'Astrophysique de Paris (UMR 7095: CNRS \& UPMC -- Sorbonne
Universit\'es), 98 bis bd Arago, F-75014 Paris, France\\ 
$^5$ Laboratoire AIM-Paris-Saclay, CEA/DSM/IRFU, CNRS, Univ. Paris VII, F-91191 Gif-sur-Yvette, France\\
$^{6}$ Department of Physics and Astronomy, The Johns Hopkins University Homewood Campus, Baltimore, MD 21218, USA\\
$^{7}$ BIPAC, Department of Physics, University of Oxford, Keble Road, Oxford OX1 3RH, UK
}

\begin{document}


\maketitle

\label{firstpage}

\begin{abstract}
We study the cross-correlation of distribution of galaxies, the Sunyaev-Zel'dovich (SZ) and X-ray power spectra of 
galaxies from current and upcoming surveys and show these to be excellent probes of the nature, i.e. extent, evolution and 
energetics, of the circumgalactic medium (CGM). 
The SZ-galaxy cross-power spectrum, especially at large multipoles, depends on the steepness of the pressure profile of 
the CGM. This property of the SZ signal can, thus, 
be used to constrain the pressure profile of the CGM. The X-ray cross power spectrum also has a similar shape. 
However, it is much more sensitive to the underlying density profile.
We forecast the detectability of the cross-correlated galaxy distribution, SZ and X-ray signals by combining 
South Pole Telescope-Dark Energy Survey (SPT-DES) and eROSITA-DES/eROSITA-LSST 
(extended ROentgen Survey with an Imaging Telescope Array-Large Synoptic Survey Telescope) surveys, 
respectively. We find that, for the SPT-DES survey, the signal-to-noise ratio (SNR) peaks at high mass and redshift with SNR
$\sim 9$ around $M_h\sim 10^{13} h^{-1} M_{\odot}$ and $z\sim 1.5\hbox{--} 2$ for flat density and temperature profiles.
The SNR peaks at $\sim 6 (12 )$ for the eROSITA-DES (eROSITA-LSST) surveys. We also perform a Fisher matrix analysis to
find the constraint on the gas fraction in the CGM 
in the presence or absence of an unknown redshift evolution of the gas fraction. 
Finally, we demonstrate that the cross-correlated SZ-galaxy and X-ray-galaxy
power spectrum can be used as powerful probes of the CGM energetics and potentially discriminate between different feedback 
models recently proposed in the literature; for example, one can distinguish a `no active galactic nuclei feedback' 
scenario from a CGM energized by `fixed-velocity hot winds' at greater than $3\sigma$.
\end{abstract}


\section{Introduction}
Galaxy formation theories predict that galaxies form by the collapse of dark matter haloes when the density crosses a critical overdensity. Baryonic matter falls into the 
gravitational potential driven by dark matter, cools down, forms stars and hence galaxies. The standard scenario predicts that the amount of baryons in a galactic halo should
approximately be a constant fraction ($\sim 16 \%$) of the total halo mass (dark matter + baryonic matter). This baryon-to-total halo mass fraction is known as the cosmic baryon fraction
(CBF). On the other hand, observations have detected only a small fraction of this CBF. More than half of the baryons of CBF are missing from the galaxies
according to soft X-ray absorption line \citep{anderson10} and emission line searches \citep{miller15} in galactic haloes. This is an important problem for all galaxy formation studies, for with 
more than 50\% of the baryon reservoir unaccounted for,  the sources and sinks for forming stars become correspondingly uncertain. Recent detection of substantial amounts of cooler gas within the halo virial radius 
fails to significantly alleviate this problem \citep{werk14}.

Although numerical simulations suggest the presence of hot coronal gas, it has been difficult to detect
this gas due to the faintness of its X-ray emission. However, there are indeed recent observations which indicate the presence of significant 
amounts of hot coronal gas extended over a large region around massive spiral galaxies \citep{anderson11, dai12, anderson13,
bogdan13a, bogdan13b}. Also, the inferred ram pressure stripping of satellite galaxies around the Milky Way Galaxy supports the 
presence of the hot halo gas with nearly flat density ($n \sim 10^{-3.5} \rm{cm}^{-3}$) out to large galacto-centric radius
\citep{putman09, putman12, gatto13}.
This gas, known as the circumgalactic medium (CGM), may account for some of the missing baryons from the galaxies. 

The CGM is the gas surrounding the central, optically visible part of the galaxy within its host dark matter halo. This is the
bridging medium that connects the interstellar medium (ISM) to the intergalactic medium (IGM). During galactic evolution, the galaxy accretes  matter from its surrounding IGM and 
also ejects some material in the form of galactic winds resulting from feedback processes like supernovae (SNe) and 
active galactic nuclei (AGN). The CGM, being 
the intermediate medium, is most affected by these processes and may contain important clues about galaxy evolution, making 
it a promising tool for the study of the processes affecting the galaxy evolution.

The presence of the hot halo gas surrounding the galaxies, in the form of the CGM, causes the Sunyaev-Zel'dovich (SZ)
distortion of cosmic microwave background radiation (CMBR). The thermal SZ (tSZ) distortion of the CMBR due to the CGM is small compared 
to  the tSZ distortion caused by the galaxy clusters \citep{planck13, singh15}. 
The detectability of the tSZ signal from a system can be enhanced by cross-correlating this signal with another signal originating
from the same source like the distribution of haloes \citep{fang12} or the gravitational lensing \citep{van14, ma14}. 
We cross-correlate the tSZ signal from the CGM with the distribution of the galaxies.
The cross-power spectrum between the SZ signal and distribution of galaxies can be thought of as the SZ-galaxy cross power spectrum.
It can be obtained by combining a high resolution a CMB survey such as SPT (South Pole Telescope) with 
an overlapping optical survey.

In addition to the SZ-effect, the hot CGM also manifests itself in X-ray emission through bremsstrahlung. Combining the X-ray observations
with optically selected galaxies can give the X-ray-galaxy cross power spectrum, which enhances the detectability of the X-ray emitting gas.
The X-ray emission from the CGM is more sensitive to the underlying gas distribution than the SZ effect and it also breaks the degeneracy
between the gas density and temperature which is present in the SZ effect. Here, we study the prospects for the 
cross-correlation of the soft X-ray emission from the CGM with
the distribution of galaxies. This can be used as an additional probe to constrain the properties of the CGM.

We also compute the X-ray-SZ cross-power spectrum for the CGM. Significant effort has been made to forecast/detect the
X-ray-SZ cross-correlation on large scales by cross-correlating the CMB maps generated by WMAP (Wilkinson Microwave Anisotropy
Probe)/{\it{Planck}} surveys with the ROSAT (R\"{o}ntgen Satellite) all sky survey \citep{diego03, hern04, hern06, hinshaw07, hajian13}. 
The X-ray-SZ cross-correlation measured for the galaxy clusters is particularly useful to constrain the cosmological 
parameters as the number of clusters strongly depends on the underlying cosmology \citep{hurier14, hurier15}.
We look into the possibility of detecting the contribution of the CGM to the X-ray-SZ cross-power spectrum with SPT and eROSITA
(extended ROentgen Survey with an Imaging Telescope Array) combination.

Simulations suggest that feedback processes are required to avoid the over-cooling problem and formation 
of excessively massive galaxies (in terms of stellar mass). A number of feedback mechanisms have been proposed that reproduce 
many observed galaxy properties despite having different implementations and physical motivations behind them. The CGM, thus 
can provide additional constraints on these simulations as the CGM properties are largely affected by the variation in the
feedback mechanism \citep{joshua15}

This paper is organized as follows: In section-\ref{sec-sze} we describe the SZ-effect. In section-\ref{sec-szg} we estimate
the cross-correlation between the SZ-effect from the hot CGM and the distribution of galaxies and forecast the detectability 
of the SZ-galaxy cross power spectrum. In section-\ref{sec-xr} we describe the X-ray emission from
the hot CGM and forecast the detectability of X-ray-galaxy cross power spectrum. In section-\ref{sec-xsz} we compute the 
X-ray-SZ cross power spectrum. In section-\ref{sec-forecast} we forecast the constraints on the CGM properties.
In section-\ref{sec-feedback} we discuss the possibility of differentiating between various feedback models.
In section-\ref{sec-conc} we conclude by summarizing our main results.

\section{SZ distortion from hot galactic halo gas}
\label{sec-sze}
For simplicity, we assume that the mass fraction of the CGM is independent of the host halo mass. Observations indicate that the fractional
 mass in the stellar component of galaxies is $\sim 0.05$ \citep{mo98, dutton10, leau10, moster10}. This corresponds to a fractional
 mass in gas, $f_{\rm gas}\,=\,0.11$, as the total fractional mass in baryons in a galaxy is $\sim 0.16$. Due to the uncertainty in the amount of the CGM, we also calculate some key results with 
 a smaller gas fraction, \fg=0.05. We assume that the gas is uniformly distributed in the galactic halo with a temperature given by the virial temperature of the halo. 
We also show the effect of different density profiles of the CGM on its cross power spectrum.

The cosmological parameters that we have used are driven by the joint analysis of  CMB anisotropies along with 
observations from the Baryon Acoustic Oscillations surveys. The resulting best fit cosmological parameters, especially 
$\sigma_8$, are in tension with those obtained from galaxy cluster and weak lensing studies. In particular, one can have 
a mismatch in the galaxy cluster counts by a factor of two due to the difference in the adopted value of $\sigma_8$.
Although, much efforts have been made to calibrate cluster masses \citep{planck15a}, crucial for doing cosmology with clusters,
the simplest explanation for the mismatch can be a remaining systematic mass bias at the galaxy cluster scales. 
Similarly, unknown systematics can also lead to the amplitude of the fluctuation spectrum inferred from  weak 
gravitational lensing to be lower than that inferred from CMB. However, this tension has been lifted to a certain
extent by the  `first' cosmological results from the Dark Energy Survey (DES, which is one of the surveys that we consider 
in our analysis), where the estimated $\sigma_8$ is consistent with the {\it{Planck}} measurement \citep{des15}.  
There are recent indications that masses estimated from velocity dispersions (for galaxy clusters) may have more robustness 
than previously envisaged \citep{rines15}. Taking positively these developments, which bring increased consensus among different 
cosmological results, we adopt the {\it{Planck}} CMB cosmological parameters \citep{planck15} as our fiducial choices. We do comment on the impact 
of cosmological parameters/degeneracies on our results  in section-\ref{sec-result}.

\subsection{Thermal SZ effect}
The inverse Compton scattering of the CMB photons by high energy electrons distorts the CMB spectrum giving rise to the tSZ
\citep{sz69}. 
The tSZ effect is represented in terms of a dimensionless parameter, known as the Compton y-parameter, defined as 
 $y= \int dl \frac{k_b T_e n_e \sigma_T}{m_e c^2}$, which for a flat density and temperature profile becomes $y= (k_b T_e n_e \sigma_TL)/(m_e c^2)$
where $\sigma_T$ is the Thomson scattering cross section, $T_e$ is the gas temperature ($T_e>>T_\gamma$),  $n_e$ is the electron density of the scattering medium, 
and $L$ is the distance travelled by the photons through the scattering medium. 
The electron density $n_e =\frac{\rho_{gas}}{\mu_e m_p}$ of the gas is determined by the condition that the total hot gas mass within the virial radius 
is a fraction $f_g=0.11$ of the total halo mass. 
For a halo of mass M at redshift $z$, the virial radius is given by
\begin{equation}
 R_{vir}=0.784  h^{-1} kpc \Bigl( \frac{M}{10^8 h^{-1}}\Bigr)^{1/3} \Bigl( \frac{\Omega_M }{\Omega_M(z)} \frac{\Delta(z)}{18 \pi^2}\Bigr)^{-1/3} \Bigl(\frac{10}{1+z} \Bigr)
\end{equation}
where $\Delta(z)=18 \pi^2+82d-39d^2$ is the critical overdensity with $d=\Omega_M(z)-1$ and $\Omega_M(z)=\Omega_M(1+z)^3/E^2(z)$,
where $E(z)=\sqrt{\Omega_{\Lambda}+\Omega_M(1+z)^3}$.

\section{SZ-galaxy cross-correlation}
\label{sec-szg}
\subsection{The angular power spectrum for tSZ}
\label{sec-tsz}
The angular Fourier transform of the Compton y-parameter (under flat sky approximation) is
\begin{equation}
  y_l = \frac{4\pi R_s}{l^2_s} \int_0 ^\infty dx x^2 y_{3D}(x)\frac{\sin(lx/l_s)}{(lx/l_s)}
  \label{eqn-yl}
\end{equation}
where $x=r/R_s$ is the dimensionless scaled radius, $R_s\equiv{R_{vir}}/{c(M,z)}$ is the scale radius, $l_s=d_A(z)/R_s$ and
$c(M,z)$ is the concentration parameter \citep{duffy08}. To calculate the SZ-effect from the  CGM, we truncate the integration in 
equation-(\ref{eqn-yl}) at $r=R_{vir}$. The 3D-radial profile $y_{3D}(x)$ is 
\begin{equation}
 y_{3D}(x)=\frac{\sigma_T}{m_e c^2} n_e(x) k_b T_e(x)
 \label{eqn-y3d}
\end{equation}
The angular power spectrum of  the tSZ effect \citep{komatsu99}
is given by
\begin{equation}
 C_l^{yy}=C_l^{yy,1h}+C_l^{yy,2h}
\end{equation}
where $C_l^{yy,1h}$ is the 1-halo or Poisson term and $C_l^{yy,2h}$ is the 2-halo or clustering term. These two terms can be written as
\begin{eqnarray}
 C_l^{yy,1h}&=& g(x_{\nu})^2\int_0 ^{z_{max}} dz \frac{dV}{dz}\int_{M_{min}} ^{M_{max}} dM \frac{dn(M,z)}{dM} |y_l(M,z)|^2 \, \nonumber\\
 C_l^{yy,2h}&=& g(x_{\nu})^2\int_0 ^{z_{max}} dz \frac{dV}{dz} P_m(k=\frac{l}{r(z)},z) W^y_l(z)^2\nonumber\\
 \label{eqn-clyy}
\end{eqnarray}
where $r(z)=(1+z)D_A$ is the comoving distance, $\frac{dV}{dz}$ is the differential comoving volume per steradian, $P_m(k,z)$ is the matter power spectrum, $b(M,z)$  is the linear
bias factor \citep{sheth99}, $\frac{dn(M,z)}{dM}$ is the differential mass function, $g(x_{\nu})=x_{\nu} \coth(x_{\nu}/2)-4 $ is the frequency dependence of the  tSZ effect with $x_{\nu}=\frac{h \nu}{k_b T_{\rm{CMB}}}$.
We compute all the power spectra in dimensionless units throughout the paper.
All calculations in this work are done in the Rayleigh-Jeans (RJ) limit ($g(x_{\nu})\rightarrow -2$). The term $W^y_l(z)$ 
present in the 2-halo term is defined as
\begin{equation}
 W^y_l(z)\equiv \int_{M_{min}} ^{M_{max}} dM \frac{dn}{dM}(M,Z) b(M,z) y_l(M,z)
 \label{eqn-wyl}
\end{equation}
We use the Sheth-Tormen mass function given by
\begin{eqnarray}
  \frac{dn}{dM}dM&=&A \sqrt{{2\alpha \nu^2 \over \pi}} \frac{\rho_m}{M^2}e^{-\alpha \nu^2 /2}
  \Bigl [-\frac{d \log\sigma}{d \log M}\Bigr ]\nonumber\\
  && \times \Bigl [1+\Bigl (\alpha \nu^2 \Bigr )^{-p} \Bigr ]dM \,,
  \label{eqn-STmassfn}
\end{eqnarray}
where $A=0.322184$, $\alpha=0.707$, $p=0.3$ \citep{sheth01} and $\nu = \frac{\delta_c}{D_g(z) \sigma(M)}$ where $\delta_c=1.68$ is the critical overdensity, $D_g(z)$ is the growth 
factor and $\sigma(M)$ is present day smoothed (with top hat filter) variance. We take $z_{max}=8$ as the upper redshift
integration limit and $M_{max}=10^{13} h^{-1} M_{\odot}$ as the upper mass integration limit. The lower mass
integration limit is set by the condition that the gas cooling time-scale is larger than the halo destruction time-scale
(explained in detail in \cite{singh15}).

\begin{figure*}
\begin{center}
 \includegraphics[width=14.0cm,angle=0.0 ]{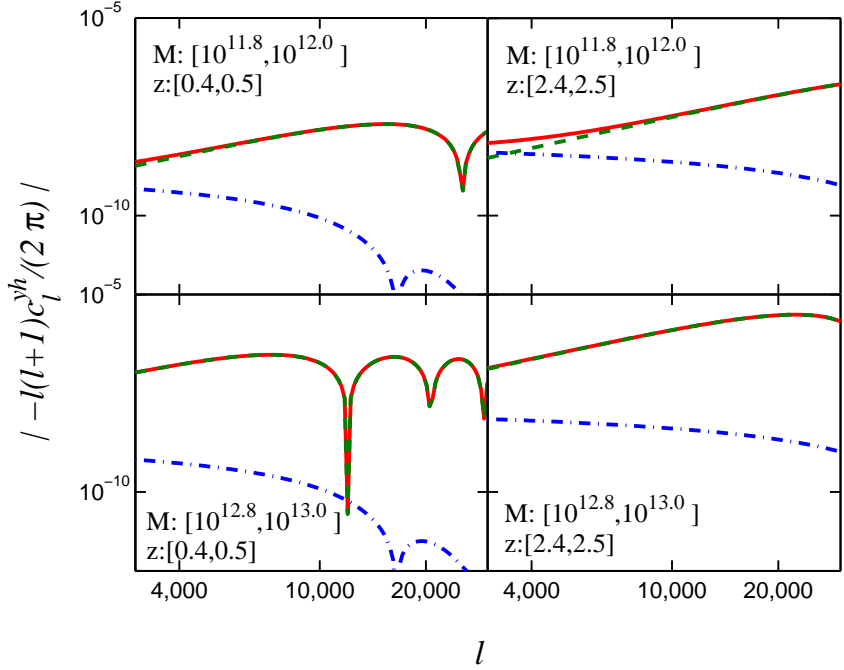}
 \caption {The cross power spectra of the SZ effect in the RJ limit and the distribution of galaxies in different mass and redshift bins. Here green (dashed) lines, blue (dot-dashed)
 lines and red (solid) lines represent the 1-halo term, 2-halo term and the total signal respectively.}
 \label{fig-cross}
\end{center}
\end{figure*}

\subsection{The angular power spectrum for the distribution of galaxies}
For simplicity, if we assume that the mass and redshift of a galaxy can be measured accurately, the probability that a given galaxy lies in the $a^{th}$ redshift bin
$z_{obs}\epsilon$ [$z^a_{obs,min},z^a_{obs,max}$] and $b^{th}$ mass bin $M_{obs}\epsilon$ [$M^b_{obs,min},M^b_{obs,max}$]
is represented by a selection function defined as \citep{oguri11, fang12}
\begin{eqnarray}
 S_{ab}(M,Z)&=&\Theta(z-z^a_{obs,min}) \Theta(z^a_{obs,max}-z)\nonumber\\
            &&\times \Theta(M-M^b_{obs,min})\Theta(M^b_{obs,max}-M)
\end{eqnarray}
where $\Theta$ is the Heaviside step function. The power spectrum for the distribution of galaxies/galactic haloes
in the $(ab)^{th}$ and $(a'b')^{th}$ bins is
\begin{equation}
 C^{hh}_{l,(ab,a'b')}= \int dz \frac{dV}{dz} P_m\Bigl(k=\frac{l}{r(z)}\Bigr) W^h_{ab}(z) W^h_{a'b'}(z)
\end{equation}
with $W^h_{ab}(z)$ defined as
\begin{equation}
 W^h_{ab}(z)\equiv \frac{1}{\bar{n}^{2D}_{ab}} \int dM \frac{dn}{dM}(M,Z) S_{ab}(M,Z) b(M,Z)
\end{equation}
where $\bar{n}^{2D}_{ab}$ is the two-dimensional angular number density of the galaxies in $(ab)^{th}$ bin and is given by
\begin{equation}
 \bar{n}^{2D}_{ab}=\int dz \frac{dV}{dz} \int dM S_{ab}(M,Z) \frac{dn}{dM}(M,Z)
\end{equation}

\subsection{SZ-galaxy cross-correlation power spectrum}
The SZ-galaxy cross power spectrum is the cross-correlation between the SZ signal and the distribution of galaxies. For the galaxies in the $(ab)^{th}$ bin, 
the SZ cross power spectrum is 
\begin{equation}
 C^{yh}_{l,(ab)}=C^{yh,1h}_{l,(ab)}+C^{yh,2h}_{l,(ab)}
\end{equation}
where $C^{yh,1h}_{l,(ab)}$ and $C^{yh,2h}_{l,(ab)}$ are the 1-halo and 2-halo terms respectively
and these terms can be written as \citep{oguri11, fang12}
\begin{equation}
 C^{yh,1h}_{l,(ab)}=\frac{g(x_{\nu})}{\bar{n}^{2D}_{ab}} \int dz \frac{dV}{dz} \int dM \frac{dn}{dM} S_{ab}(M,Z) y_l(M,Z) 
 \label{eqn-clyh1h}
\end{equation}

\begin{equation}
 C^{yh,2h}_{l,(ab)}=g(x_{\nu}) \int dz \frac{dV}{dz} P_m\Bigl(k=\frac{l}{r(z)}\Bigr) W^h_{ab}(z) W^y_l(z)
 \label{eqn-clyh2h}
\end{equation}
Due to the presence of $S_{ab}(M,Z)$ in Eq.\ref{eqn-clyh1h} and \ref{eqn-clyh2h}, only the galaxies lying in the $a^{th}$ redshift
bin contribute to the SZ cross power spectrum. For the 2-halo term, even the galaxies lying outside the $b^{th}$ mass bin contribute because of the 
the presence of $W_l ^y(z)$ in Eq.\ref{eqn-clyh2h} (see Eq.\ref{eqn-wyl}).

In Figure-(\ref{fig-cross}) we show the SZ-galaxy cross power spectrum for a few mass and redshift bin combinations.
The galaxies are binned in mass (total halo mass) and redshift with $\Delta \log(M)=0.2$ and $\Delta z=0.1$ respectively. The mean
redshift increases from left panel to right panel and the mean halo mass increases from top panel to bottom panel. Most 
of the contribution to the total cross power spectrum comes from the 1-halo term. 

\subsection{CGM density profile and the cross-power spectrum}
\label{sec-flct}
\begin{figure}
\begin{center}
 \includegraphics[width=9.0cm,angle=0.0 ]{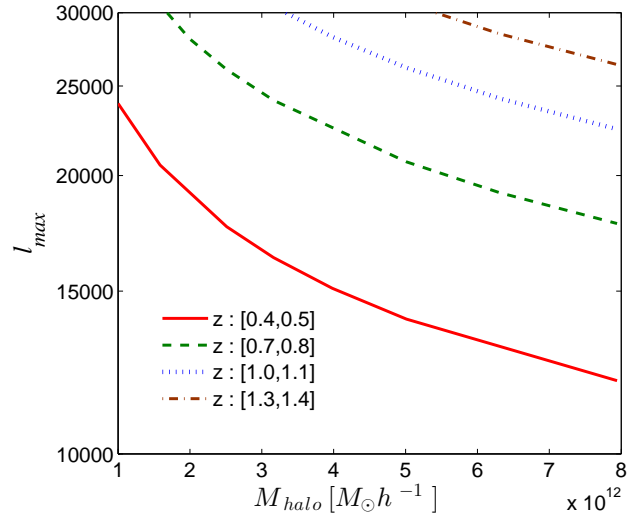}
 \caption {$l_{max}$ is shown as a function of the mean halo mass for redshift bins $z:[0.4\hbox{--}0.5]$ (red solid line)
 , $z:[0.7\hbox{--}0.8]$ (green dashed line), $z:[1.0\hbox{--}1.1]$ (blue dotted line) and $z:[1.3\hbox{--}1.4]$ (brown
 dot-dashed line).}
 \label{fig-lmax}
\end{center}
\end{figure}

For a flat pressure profile for the CGM, the SZ-galaxy cross power spectrum shows oscillations if $l>l_{max}$, where $l_{max}$ depends on the 
mass and redshift of the galaxy. In Figure-(\ref{fig-lmax}) we show $l_{max}$ as a function of the 
mean halo mass for different redshift bins. The oscillations begin when the multipole $l$ corresponds to an angular size
$\sim \frac{2}{3} \times$ virial radius of the galaxy. 
The reason for these oscillations is the truncation of the signal at $R_{vir}$.

\begin{figure}
\begin{center}
 \includegraphics[width=9.0cm,angle=0.0 ]{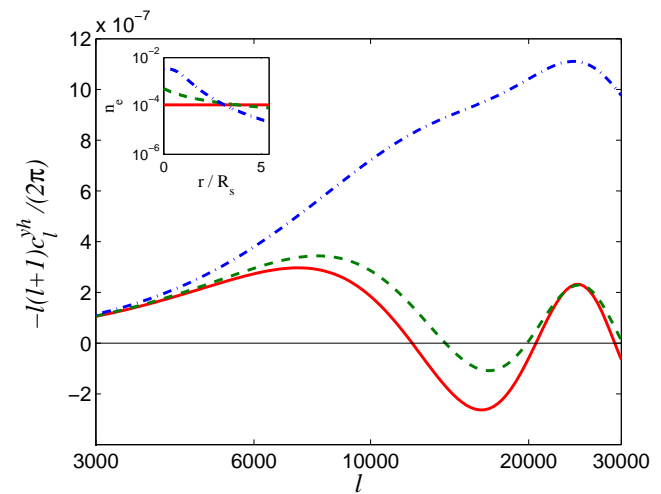}
 \caption {The cross power spectra of the SZ effect (in RJ limit) and the distribution of the galaxies for $M_h:[10^{12.8},10^{13.0}] (h^{-1}M_{\odot})$
 and $z:[0.4,0.5]$ for the density profile-(a) (red solid line), profile-(b) (green dashed line) and profile-(c) (blue dot-dashed line) i.e. 
 \gmag$=0,\,1\,\&\,3$ respectively. In the inset, we show the corresponding density profiles.}
 \label{fig-prf}
\end{center}
\end{figure}

The shape of the SZ-galaxy cross power spectrum is sensitive to the pressure profile of the CGM.
Since we have fixed the temperature of the CGM, the density profile of the CGM can be constrained using the 
cross-correlated SZ-galaxy power spectrum.
In order to estimate the effect of different density profiles on the SZ cross power spectrum, we parameterize the density
profile by \gmag, defined by $\rho_{gas} \propto [1+(r/R_s)^{\gamma_{\rm{gas}}}]^{-1}$
In Figure-(\ref{fig-prf}) we show the total SZ cross-power spectrum for mass bins:$[10^{12.8},10^{13.0}] (h^{-1}M_{\odot})$ and 
redshift bins:$[0.4,0.5]$ for the following three density profiles,\\\\
\noindent
Profile-(a): \gmag=0 $\Longrightarrow$ a flat density profile.\\
\noindent
Profile-(b): \gmag=1 $\Longrightarrow$ $\rho_{gas} \propto [1+(r/R_s)]^{-1}$\\
\noindent
Profile-(c): \gmag=3 $\Longrightarrow$ $\rho_{gas} \propto [1+(r/R_s)^3]^{-1}$\\\\
From profiles-(a) to (c), the density becomes steeper and more centrally concentrated. 
Above density profiles are similar to the $\beta$-model (with $\beta = 2/3$) with a central core followed by a gradual decrease 
in the density. This choice of the density profile is inspired by the observation of nearly flat distribution of the hot halo gas
in the Milky Way Galaxy \citep{putman09, putman12, gatto13} and the simulation results of \cite{brun15} and \cite{joshua15},
which predict a flatter distribution of the gas compared to the Navarro-Frenk-White (NFW) profile in galaxies due to the 
presence of feedback processes.
In all 3 cases the SZ signal is truncated
at the virial radius. As the density profile becomes steeper from (a) to (c) the value of $l_{max}$ shifts from 12000 to 14000 whereas 
there are no oscillations in the case of profile-(c). This shift occurs since with the steepening of the density profile, the 
pressure at the virial radius decreases.

In Figure-(\ref{fig-prf}) at small $l$-values ($\sim 3000$), the SZ cross power spectrum for the three density profiles are almost identical
but there is a significant difference between the profiles near $l\sim 10^4$. This is because the steepening of the density profile
increases the power at small angular scales or large $l$ values. The value of $l$ where the shape of the cross-power spectrum
is significantly different for different density profiles depends on the mean mass and redshift of the bin in a similar way as for $l_{max}$. 
Therefore, the shape of the SZ-galaxy cross-power spectrum at these $l$-values can be used to determine the slope of the 
density profile of the CGM. However, the use of 
this method is limited by the resolution of the CMB survey. Presently the SPT survey has the best resolution and it goes up to
$l\approx10^4$. High mass and low redshift galaxies are better choices for this purpose as the SZ power spectrum for different
profiles is distinguishable at $l<10^4$ for these galaxies.

\subsection{Detectability of the SZ-galaxy cross-correlation signal}
\label{sec-dectsz}
\subsubsection{Cumulative signal-to-noise ratio} 
\label{sec-snrsz}
Given a survey one can estimate the detectability of the SZ-galaxy cross-correlation signal for galaxies binned according to their masses and redshift using the cumulative signal-to-noise ratio
(SNR) defined as
\begin{equation}
 \Bigl( \frac{S}{N}\Bigr)^2 = \Sigma_{l l'} C^{yh}_{l,(ab)} (M^{yh}_{ll',(ab)})^{-1} C^{yh}_{l',(ab)} \delta_{ll'}
\end{equation}
where $M^{yh}_{ll',(ab)}$ is the covariance matrix which represents the uncertainty in the measurement of $C^{yh}_{l,(ab)}$. The 
covariance matrix is given by
\begin{equation}
 M^{yh}_{ll',(ab,a'b')} = \frac{\delta_{ll'}}{f_{sky}(2l+1)\Delta l} \times \Bigl[\hat{C}^{yy}_{l} \hat{C}^{hh}_{l,(ab,a'b')}+
 \hat{C}^{yh}_{l,(ab)} \hat{C}^{yh}_{l,(a'b')}\Bigr]
 \label{mll}
\end{equation}
where $f_{sky}$ is the fractional sky coverage for a survey, $\Delta l$ is the $l$-bin size used to calculate the power spectrum and $\hat{C}_l$'$s$ represent the power
spectrum including the noise contribution (i.e. $C^i_l+N^i_l$) where i stands for $yy$, $hh$ and $yh$. For simplicity, we neglect
the non-Gaussian contribution for the $l$-range considered: this is particularly true for the massive galaxies.
Note that the dominant contribution to the noise in the SZ-galaxy as well as X-ray-galaxy cross-power spectrum comes from the instrumental 
noise. Thus, the inclusion of clusters has negligible effect on the error calculations and hence the SNR. Therefore, we neglect this contribution in 
our calculation.

The instrumental noise simply adds to the SZ power spectrum and for a given CMB survey, it is given by 
\begin{equation}
 N^{yy,(CMB)}_l=\frac{1}{\Sigma_k w_k s^2_k B^2_{kl}}
 \label{nscmb}
\end{equation}
where the summation in Eq.\ref{nscmb} is over the different frequency channels, $w=(\sigma_T \theta_{\rm{FWHM}}/T_{\rm{CMB}})^{-2}$ where $\sigma_T$ is the rms instrumental noise
per pixel, $\theta_{\rm{FWHM}}$ is the full width half-maximum of the beam, $B^2_l=\exp\Bigl[-l(l+1)\theta^2_{\rm{FWHM}}/(8ln2)\Bigr]$ is the fourier transform of beam
profile and $s=-g(x_{\nu})/2$ is to rescale the result in RJ limit. 
The shot noise in the galaxy distribution power spectrum is
\begin{equation}
 N^{hh,(g)}_{(ab,a'b')}=\frac{1}{n^{2D}_{ab}}\delta_{aa'}\delta_{bb'}
 \label{nshalo}
\end{equation}

For the cross-power spectrum $N^{yh}_l=0$ i.e. $\hat{C}^{yh}_l=C^{yh}_l$ as the distribution of galaxies is not 
correlated with the instrumental noise in the CMB surveys.

\begin{table}
\caption{Specifications of surveys}
\centering 
\begin{tabular}{l l l l l l}
\hline
 Survey & \hspace{2 mm}$\Omega_s$ & $f_{sky}$ & \hspace{4 mm}$\nu$ & $\theta_{\rm{FWHM}}$ & $\sigma_T$ \\
 & (deg$^2$) &  & $(GHz)$ & $(arcmin)$ & $(\mu K)$ \\
 
 \hline
  &  &  & 95  & 1.6 & 26.3  \\[-0.5ex]
  SPT &  2500  &  6\%   & 150 & 1.1 & 16.4  \\[-0.5ex]
    &   &     & 220 & 1.0 & 85 \\[-0.5ex]
  \hline 
  \hline\\
 \end{tabular}

 \begin{tabular}{l l l l}
 \hline
 Survey  & \hspace{8 mm}$\Omega_s$ & \hspace{12 mm}$f_{sky}$ & \hspace{12 mm}$z$ range \\
&  \hspace{6 mm}(deg$^2$) &  &  \\

  \hline
 DES & \hspace{6 mm}5000 & \hspace{12 mm}12\% & \hspace{12 mm}0.1-1.4  \\[-0.5ex]\\
 LSST & \hspace{5 mm}20000 & \hspace{12 mm}48\% & \hspace{12 mm}0.1-1.4  \\[-0.5ex]
 \hline
 \hline\\
 \end{tabular}

 \begin{tabular}{l l l l l}
 \hline
 Survey & \hspace{2 mm}$\Omega_s$ & $f_{sky}$ & $\theta_{\rm{FWHM}}$ & Exposure time \\
  &   (deg$^2$)   &           & $(arcsec)$ & \hspace{5 mm}($ks$) \\
 \hline
 eROSITA & All sky & 100\% & \hspace{5 mm}20 & \hspace{5 mm}2 \\[-0.5ex]
   \hline
 \end{tabular}

 \label{tab1}
\end{table}

\begin{figure}
\begin{center}
 \includegraphics[width=9.50cm,angle=0.0 ]{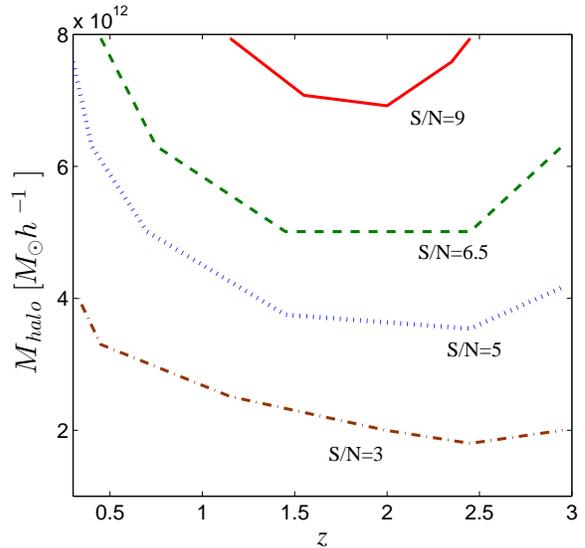}
 \caption {Contours of the cumulative signal-to-noise (SNR) for the measurement of the cross-power spectrum of the SZ effect with the distribution of 
 galaxies for the combined SPT-DES like surveys. 
 The red (solid), green (dashed), blue (dotted) and brown (dot-dashed) lines represent the SNR 9, 6.5, 5 and 3 respectively.
 The upper redshift limit plotted here is more than the highest redshift probed by DES.}
 \label{fig-contourf}
\end{center}
\end{figure}

\subsubsection{CMB survey}
To detect the SZ-galaxy cross-correlation signal, we need a galaxy survey and a CMB survey with overlapping sky coverages. Since the SZ
signal from the CGM becomes non-negligible compared to other contribution to CMB distortion at large $l\hbox{--}$values 
( at $l \gtrsim 3000$ ), we consider the SPT survey for this work. Presently, the optical survey which overlaps with the SPT survey, is the DES. 
The specifications of these surveys are given in Table-(\ref{tab1}).

Note that, the dominant contribution to the covariance in Eq.\ref{mll} comes from 
$N^{yy,(CMB)}_l / n^{2D}_{ab}$ for $l>3000$. The contribution from other terms ($C^{yy}_l C^{hh}_l$, $C^{hh}_l N^{yy,(CMB)}_l$, 
$C^{yy}_l / n^{2D}_{ab}$ and $C^{yh}_l C^{yh}_l$) is negligible compared to this term. For example, for the haloes in the mass bin
$M_h:[10^{12.8},10^{13.0}] (h^{-1}M_{\odot})$ and redshift bin $z:[0.4,0.5]$, the instrumental noise is nearly three to four orders of 
magnitude larger than $C^{yh}_l$.\\

In figure-(\ref{fig-contourf}) we show the contours of cumulative SNR of the SZ-galaxy cross power spectrum of the CGM for the flat density
profile. The galaxies are binned in mass and redshift with $\Delta \log(M)=0.2$ and $\Delta z=0.1$ respectively.
Here we take $l_{min}=3000$. Note that the DES survey goes only up to $z\sim1.4$ whereas in 
Figure-(\ref{fig-contourf}) we have shown the results up to $z\sim3$ to show the decreasing contribution from high redshift
haloes. The SZ signal increases with increasing halo mass for a given redshift as the amount of gas causing the SZ-effect
increases with increasing halo mass. This results in higher SNR for higher mass galaxies compared to the low mass haloes in the same 
redshift bin. For a given halo mass, the SNR first increases with increasing redshift, reaches a maximum value and then 
decreases. As a combined effect there is an optimum spot with SNR $\sim 9$ at the high mass end of galaxies, at redshift around $z\sim 1.5\hbox{--}2$.
Even for low mass galaxies ($M_h \sim 2\hbox{--}4 \times 10^{12} h^{-1} M_{\odot}$) the SNR is $\sim 3$ in the redshift range
covered by the DES survey. This makes the SPT-DES survey a useful tool to study the CGM in the low mass galaxies which is otherwise difficult
to detect at higher redshifts. 

In the calculation of the SNR in this section as well as the constraints on the model parameters from the SZ-galaxy cross-power spectrum in
section-\ref{sec-result} and \ref{sec-feedback}, we neglect the contamination from other astrophysical sources (kinetic
SZ (kSZ), Cosmic infrared background (CIB), point sources etc.). However, the tSZ signal has a distinct frequency dependence
and the relative contribution from other sources can be minimized using multi-frequency data from surveys like
SPT.

\begin{figure*}
\begin{center}
 \includegraphics[width=14.0cm,angle=0.0 ]{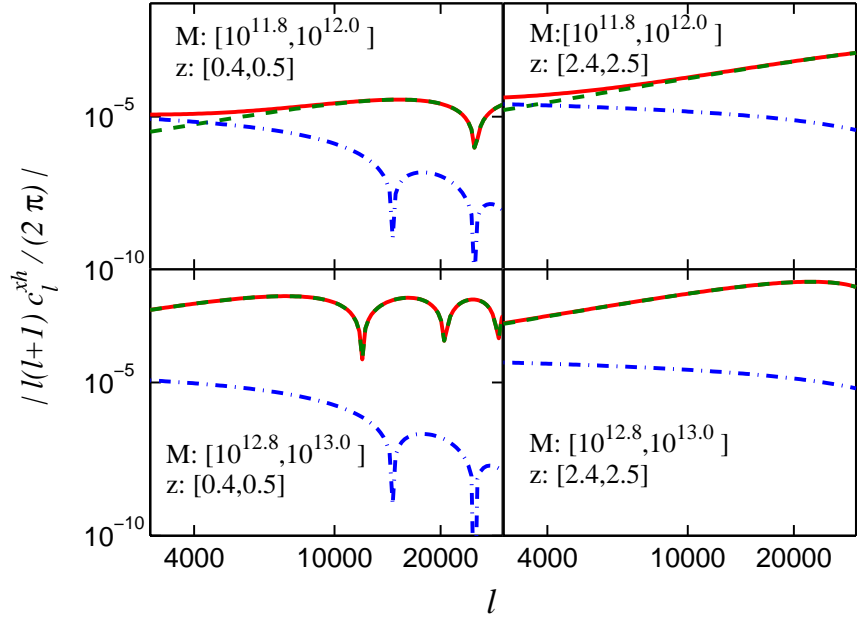}
 \caption {The cross power spectra of the X-ray emission from the CGM and the distribution of galaxies in different mass and redshift bins. Here green (dashed) lines, blue (dot-dashed)
 lines and red (solid) lines represent the 1-halo term, 2-halo term and the total signal respectively.}
 \label{fig-subxh}
\end{center}
\end{figure*}

\section{X-ray-galaxy cross-correlation}
\label{sec-xr}
The hot CGM causing the SZ-effect also emits in the soft X-rays. This X-ray emission from the galaxies can be cross-correlated with the galaxy distribution 
to increase its detectability. The X-ray surface brightness due to the presence of the hot gas in a direction $\theta$ (eg. \cite{cheng03}) 
is given by 
\begin{equation}
 S(\theta)\approx \frac{1}{4 \pi (1+z)^4} \int n^2_e \Lambda(T,Z) d\chi
\end{equation}
where $\Lambda(T,Z)$ is the cooling function which depends on the metallicity and temperature of the gas. 
We assume that the halo gas has a metallicity $\sim 0.1\,Z_{\odot}$ and use the cooling function from 
\cite{dopita93}. For the metallicity and mass range of interest, we calculate the emission from the CGM in the soft X-ray band ($0.5$-$2.0$ $\rm{keV}$) assuming that
the fraction of total energy in soft X-ray band is $\sim (\exp^{-(\frac{E_1}{k_b T})}-\exp^{-(\frac{E_2}{k_b T})})$ , where 
E$_1$ is the lower and E$_2$ is the higher energy limit of the soft X-ray band. The mass and redshift range considered
here corresponds to the temperature range $\sim 10^{6}$-$10^7$ K. Therefore, these galaxies lie near the lower energy
limit of the soft X-ray band used for this study.

The fluctuations in the X-ray background in the direction $\theta$ are
\begin{equation}
 s(\theta)=\frac{S(\theta)}{<S_{\rm{SXRB}}>}-1
\end{equation}
where $<S_{\rm{SXRB}}>$ is the mean surface brightness of the soft X-ray background (SXRB) and it includes all possible sources 
of SXRB (see \cite{merloni12}).

\subsection{The X-ray angular power spectrum}
In  analogy with the SZ-effect, the angular Fourier transform of fluctuations in the SXRB (for $l>0$) is given by
\begin{equation}
 s_l= \frac{4\pi R_s}{l^2_s} \int_0 ^\infty dx\, x^2 s_{3D}(x)\frac{\sin(lx/l_s)}{(lx/l_s)}
  \label{eqn-sl}
\end{equation}
where the $3D$-radial profile $s_{3D}(x)$ is 
\begin{equation}
 s_{3D}(x)=\frac{1}{4 \pi (1+z)^4} \frac{n^2_e(x) \Lambda(T,Z)}{<S_{\rm{{SXRB}}}>}
\end{equation}
The X-ray power spectrum can be obtained by replacing $g(x_{\nu})y_l$ by $s_l$ in Eq-(\ref{eqn-clyy}).
\begin{eqnarray}
 C_l^{xx,1h}&=& \int_0 ^{z_{max}} dz \frac{dV}{dz}\int_{M_{min}} ^{M_{max}} dM \frac{dn(M,z)}{dM} |s_l(M,z)|^2 \, \nonumber\\
 C_l^{xx,2h}&=& \int_0 ^{z_{max}} dz \frac{dV}{dz} P_m(k=\frac{l}{r(z)},z) W^x_l(z)^2\nonumber\\
 \label{eqn-clxx}
\end{eqnarray}
with $W^x_l(z)$ defined as
\begin{equation}
 W^x_l(z)\equiv \int_{M_{min}} ^{M_{max}} dM \frac{dn}{dM}(M,Z) b(M,z) s_l(M,z)
 \label{eqn-wxl}
\end{equation}
Similarly, the X-ray-galaxy cross-power spectrum is given by
\begin{eqnarray}
 C^{xh,1h}_{l,(ab)}&=&\frac{1}{\bar{n}^{2D}_{ab}} \int dz \frac{dV}{dz} \int dM \frac{dn}{dM} S_{ab}(M,Z) s_l(M,Z) \, \nonumber\\ 
 C^{xh,2h}_{l,(ab)}&=& \int dz \frac{dV}{dz} P_m\Bigl(k=\frac{l}{r(z)}\Bigr) W^h_{ab}(z) W^x_l(z)
 \label{eqn-clxh}
\end{eqnarray}

To calculate $<S_{\rm{SXRB}}>$ we use the SXRB counts expected to be observed by the eROSITA mission \citep{merloni12} 
assuming a conversion factor $1\, \rm{count/sec} \sim 10^{-11} erg\, sec^{-1} cm^{-2}$, which is approximately 
the conversion factor for the ROSAT all sky survey. Note that using a constant conversion factor underestimates the power
at low redshifts (z$<$1 for massive galaxies) and overestimates the power at high redshifts (z$>$1). Therefore, this
simplified approach gives only an order of magnitude estimate of the X-ray power spectrum. However, this does not affect
the estimate of the uncertainty on the model parameters using X-ray-galaxy cross-power spectrum as the main contribution
to the constraints comes from the massive galaxies near z$\sim$1. In addition, these constraints saturate fast once the 
information from two or more appropriate mass-redshift bins is combined together as shown in section-\ref{sec-result}.

In Fig-(\ref{fig-subxh}) we show the X-ray-galaxy cross power spectrum for a few mass and redshift bins. For the flat density and temperature 
profiles, the shape of the X-ray-galaxy cross-correlation power spectrum is analogous to the shape of the SZ-galaxy cross power spectrum and has the same
value of $l_{max}$. The contribution of the 2-halo term is negligible compared to the 1-halo term beyond $l=10^4$ for all mass and 
redshift bin combinations considered here.

\begin{figure}
\begin{center}
 \includegraphics[width=9.0cm,angle=0.0 ]{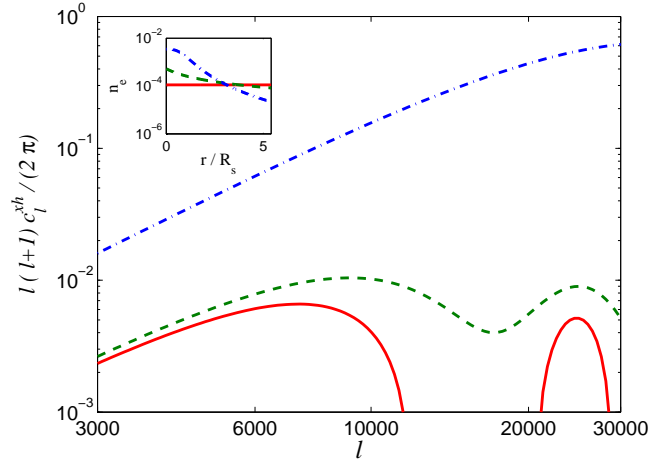}
 \caption {The X-ray-galaxy cross power spectra for $M_h:[10^{12.8},10^{13.0}] (h^{-1}M_{\odot})$
 and $z:[0.4,0.5]$ for the density profile-(a) (red solid line), profile-(b) (green dashed line) and profile-(c) (blue dot-dashed line).
 In the inset we show the corresponding density profiles.}
 \label{fig-xprf}
\end{center}
\end{figure}

In Figure-(\ref{fig-xprf}) we show the X-ray-galaxy cross power spectrum for $M_h:[10^{12.8},10^{13.0}] (h^{-1}M_{\odot})$
 and $z:[0.4,0.5]$ for the density profiles described in section-\ref{sec-flct}. Note that for all these profiles we have
 assumed the CGM to be at the virial temperature and we truncate the signal at the virial radius. The X-ray-galaxy cross-correlation signal increases
 rapidly compared to the SZ-galaxy cross-correlation signal with the steepening of the density profile. This is due the fact that the X-ray emission
 is proportional to $n^2_e$ and is more sensitive to the gas density profile compared to the SZ-effect which is proportional
 to $n_e$. The difference between the profiles is now significant even at smaller $l$-values ($\sim 3000$).

\subsection{Detectability of the X-ray-galaxy cross-correlation signal}
As mentioned earlier in section-\ref{sec-snrsz}, the cumulative signal-to-noise ratio provides an efficient
way of estimating the detectability of a signal for a given survey. The cumulative SNR of the X-ray-galaxy cross power spectrum in $ab^{th}$ bin is
\begin{equation}
 \Bigl( \frac{S}{N}\Bigr)^2 = \Sigma_{ll'} C^{x}_{l,(ab)} (M^{xh}_{ll',(ab)})^{-1} C^{xh}_{l',(ab)} \delta_{ll'}
\end{equation}
where the covariance matrix $M^{xh}_{ll,(ab)}$ for the X-ray-galaxy cross-correlation is given by
\begin{equation}
  M^{xh}_{ll',(ab,a'b')} = \frac{\delta_{ll'}}{f_{sky}(2l+1)\Delta l} \times \Bigl[\hat{C}^{xx}_{l} \hat{C}^{hh}_{l,(ab,a'b')}+
 \hat{C}^{xh}_{l,(ab)} \hat{C}^{xh}_{l,(a'b')}\Bigr]
 \label{eq-noisex}
\end{equation}
with $\hat{C}_l=C^i_l+N^i_l$. Assuming that the noise in X-rays is dominated by the shot noise, the noise in the X-ray power 
spectrum is
\begin{equation}
 N^{xx}_l=\frac{1}{N_{bg}} \exp\Bigl(\frac{l(l+1)\theta^2_{\rm{FWHM}}}{8ln2}\Bigr)
\end{equation}
where $N_{bg}$ is the total number of the soft X-ray photons collected/steradian by the X-ray telescope and $\theta_{\rm{FWHM}}$ is 
the full width half maximum of the beam.

In practice some of the X-ray background is produced by the X-ray emission of AGNs, X-ray binaries, SNe remnants amongst
which AGNs are dominant. AGNs are clustered
with galaxies and would thus contribute to the X-ray-galaxy correlation. This could however be mitigated by first
removing the fraction of X-ray AGNs which are above the detection threshold of the X-ray survey. In addition, AGNs
have a harder X-ray spectrum than the diffuse circumgalactic X-ray gas. The photon energy dependence of the cross-correlation
signal can thus also be used to separate the contribution from AGNs. In addition, the signal from AGNs would be produced
by a 1-source term and a clustering term with an angular dependence determined by the point spread function (PSF) of the X-ray instrument and the
correlation function of the AGNs. This specific angular dependence can be used to disentangle the contribution from the AGNs and
from the circumgalactic gas. While a detailed analysis which incorporates these mitigating techniques is beyond the scope
of this paper, we make the optimistic assumption that the X-ray noise is not correlated with the distribution of
galaxies, i.e. $N_l^{xh}=0$ in Eq.\ref{eq-noisex}.

\subsection{X-ray survey}
Since we are interested in the X-ray signal from the galaxies, we need an X-ray survey with a small beam size (high 
resolution) and large sky coverage. We consider the eROSITA survey for this purpose. The eROSITA
is a future mission expected to be launced in 2016 (see \cite{merloni12} for the details of this mission). 
The specifications of this mission are given in 
Table-(\ref{tab1}). The total background expected in the soft band of the eROSITA is $\sim 2\times10^{-3}$ $\rm {counts\,sec^{-1}arcmin^{-2}}$
\citep{merloni12}.
To calculate the X-ray-galaxy cross-power spectrum, we consider the combination of the eROSITA-DES and eROSITA-LSST (Large Synoptic Survey Telescope) surveys. 

\begin{figure}
\begin{center}
 \includegraphics[width=8.7cm,angle=0.0 ]{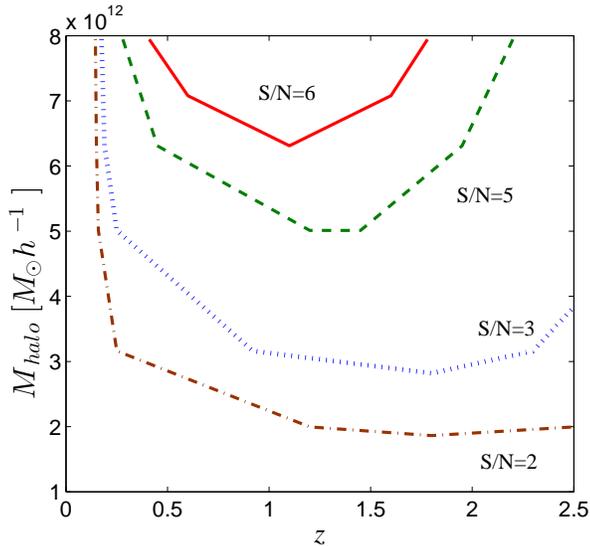}
 \caption {Contours of the cumulative signal-to-noise (SNR) for the measurement of the cross-power spectrum of the X-ray emission from the CGM with the distribution of 
 galaxies for the combined eROSITA-DES-like survey. 
 The red (solid), green (dashed), blue (dotted) and brown (dot-dashed) lines represent the SNR 6, 5, 3 and 2 respectively.
 Note that the upper redshift limit plotted here is more than the highest redshift probed by DES.}
 \label{fig-contourx}
\end{center}
\end{figure}

In Fig-(\ref{fig-contourx}) we show the contours of the cumulative SNR for the X-ray-galaxy cross power spectrum for the eROSITA-DES combination. 
Similar to the SZ-galaxy cross power spectrum, the galaxies are binned in mass and redshift with $\Delta \log(M)=0.2$ and $\Delta z=0.1$ respectively.
Again the high mass galaxies have larger SNR due to their larger gas reservoir compared to the low mass galaxies and the
SNR increases with increasing redshift, becomes maximum and then decreases with further increase in redshift. The difference
between the X-ray-galaxy and SZ-galaxy cross power spectra is that the X-ray cross-power spectrum peaks at relatively smaller redshift compared to the SZ
cross-power spectrum due to the fact that the observed X-ray surface brightness decreases rapidly with increasing redshift. Also
the contribution from the low mass galaxies at high redshifts is more than that of the SZ-galaxy cross power spectrum. This is essentially due to the 
much better resolution of the eROSITA ($\sim 20 ''$) compared to the resolution of the SPT ($\sim 1 '$).
The SNR peaks for the high mass and intermediate redshift galaxies $(z\sim 1)$. For the most massive galaxies ($M_{h}=[10^{12.8},10^{13.0}] h^{-1}M_{\odot}$)  
at redshift $z \sim 1$, the SNR is $\sim 7$. For the low mass galaxies ($ \lesssim 10^{12} h^{-1}M_{\odot}$), the SNR is low 
($<2$) at all redshifts. Massive galaxies have significant SNR ($\sim 6$) even at redshifts $<0.5$, hence the X-ray-galaxy 
cross-power spectrum can be used to study the CGM of these systems.

The sky coverage of the LSST is 4 times the sky coverage of the DES. Since the cumulative SNR 
$\propto \sqrt{f_{sky}}$, SNR for the eROSITA-LSST combination will be twice the SNR of the eROSITA-DES combination. 
Therefore, for the assumed CGM properties, the eROSITA-LSST survey will be able to detect the X-ray-galaxy cross-correlational signal
from the galaxies with a peak SNR $\sim 14$. Note that the estimates of the SNR for both the SZ-galaxy and X-ray-galaxy
power spectra depend on the size of the mass and redshift bins. Therefore, increasing or decreasing the size of mass and/or 
redshift bin also changes the detectability of these signals accordingly.

\begin{figure}
\begin{center}
 \includegraphics[width=9cm,angle=0.0 ]{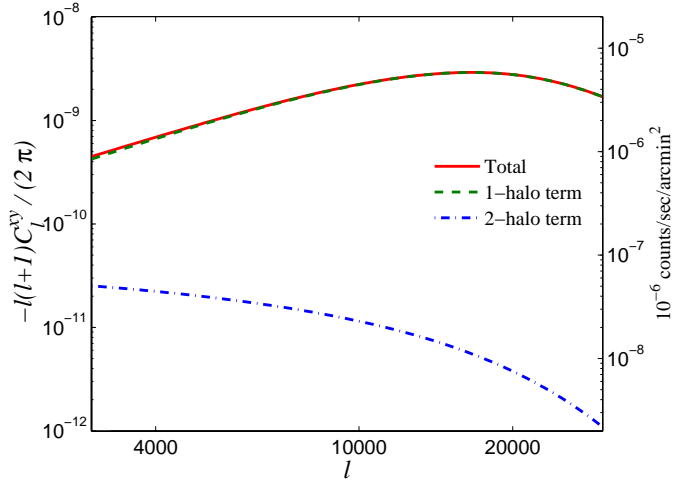}
 \caption {The cross power spectra of the SZ effect in RJ limit and soft X-ray emission from CGM. Here green (dashed) lines, blue (dot-dashed)
 lines and red (solid) lines represent the 1-halo term, 2-halo term and the total signal respectively.}
 \label{fig-xsz}
\end{center}
\end{figure}

\begin{table}
\caption{Fiducial values and priors on the parameters}
\centering 
\begin{tabular}{l c c c c }
 \hline \hline
 Parameter & Fiducial value & \pra & \prb  \\[0.5ex]
 \hline
$\sigma_8$ & 0.831 & 0.013 & 0.013 \\
$\Omega_M$ & 0.3156 & 0.0091 &0.0091 \\
$n_s$ & 0.9645 & 0.0049 & 0.0049 \\
$f_{\rm gas}$ & 0.11 & - & - \\
$f_{\rm Temp}$ & 1.0 & -  & 0.25 \\
$\alpha_{\rm gas}$ & 0.0 & - & - \\
\hline
 \end{tabular}
\label{tab_fiducial}
\end{table}

\section{X-ray-SZ cross-correlation}
\label{sec-xsz}
The SZ effect and X-ray emission have different dependences on the gas density and temperature. The SZ effect is proportional to 
$n_e T_e$ whereas the X-ray emission scales approximately as $n^2_e \sqrt{T_e}$. Combining the two can improve the constraints
on the gas physics parameters. The X-ray-SZ cross power spectrum is given by 
\begin{equation}
 C_l^{xy}=C_l^{xy,1h}+C_l^{xy,2h}
\end{equation}
where
\begin{eqnarray}
 C_l^{xy,1h}&=& g(x_{\nu})\int_0 ^{z_{max}} dz \frac{dV}{dz}\int_{M_{min}} ^{M_{max}} dM \frac{dn(M,z)}{dM}[s_l(M,z) \nonumber\\
  &&\hspace{50 mm}\times y_l(M,z)] \, \nonumber\\
 C_l^{xy,2h}&=& g(x_{\nu})\int_0 ^{z_{max}} dz \frac{dV}{dz} P_m(k=\frac{l}{r(z)},z) [W^x_l(z)\nonumber\\
       &&\hspace{50 mm}\times W^y_l(z)]
 \label{eqn-clxy}
\end{eqnarray}

In Figure-(\ref{fig-xsz}) we show the X-ray-SZ cross power spectrum for the flat density profile and the mass and redshift range specified in 
section-\ref{sec-tsz} in dimensionless as well as $10^{-6}\rm {counts\,sec^{-1}arcmin^{-2}}$ units.
The cross-power spectrum peaks at $l\approx 15000$ for galactic haloes whereas for galaxy clusters it peaks at
$l\approx3000$ (see Figure-1 of \cite{hurier14}). This difference is mainly because the galaxies are smaller objects than
the clusters and therefore, the X-ray-SZ cross power spectrum for galaxies peaks at smaller angular scales or larger $l$-values.
Recently, \cite{hurier15} detected the total X-ray-SZ cross-power spectrum at 28$\sigma$ level with ROSAT and 
{\it{Planck}} all sky surveys. For the $l$-range of interest for the CGM, we show the SPT-eROSITA combination. 
 However, due to the weak signal compared to the noise 
for SPT-eROSITA surveys (SNR$\sim 0.65$), it is not possible to detect the X-ray-SZ cross-correlation signal from the CGM.

\section{Forecasting of CGM constraints}
\label{sec-forecast}
We now use the Fisher matrix formalism to forecast the expected constraints for different survey combinations.
Since the cosmological parameters are well constrained by {\it{Planck}}, our main focus is to constrain the astrophysical
parameters related to the gas physics. 
The parameters considered for this work are
\begin{equation}
 \{[\sigma_8, \Omega_M, n_s],[ f_{\rm gas},  f_{\rm Temp}, \alpha_{\rm gas}] \} \;\; ,
 \label{fischer}
\end{equation}
where the parameters in the first bracket are cosmological parameters and in the second bracket are astrophysical
parameters which depend on baryonic physics. Note that we have assumed the flat-$\Lambda$-cold dark matter cosmology.
The fiducial values of these parameters are given in Table-(\ref{tab_fiducial}). Here $f_{\rm gas}$ is the redshift independent gas fraction, $f_{\rm Temp} = \frac{T}{T_{\rm vir}}$, i.e.,  
the ratio of the temperature of the gas to the virial temperature of the gas in the halo 
and $\alpha_{\rm gas}$ represents any possible evolution of the gas defined through
$f_{\rm gas}(z)=f_{\rm gas}[E(z)]^{\alpha_{\rm gas}}$. Our fiducial model assumes no redshift evolution of the gas fraction
i.e. \alpg=0.

\begin{figure*}
\begin{center}
 \includegraphics[width=8cm,angle=0.0 ]{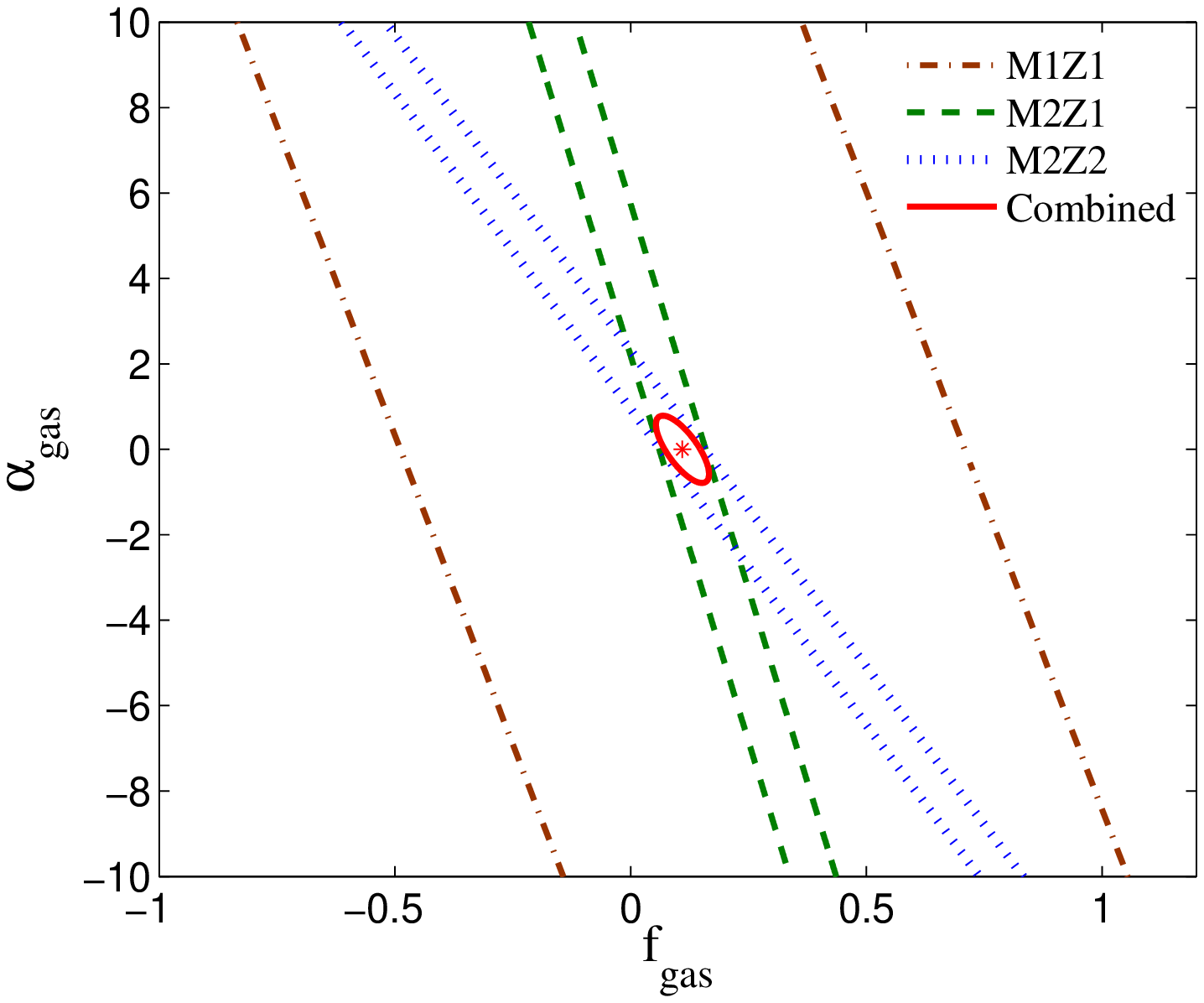}
 \includegraphics[width=8cm,angle=0.0 ]{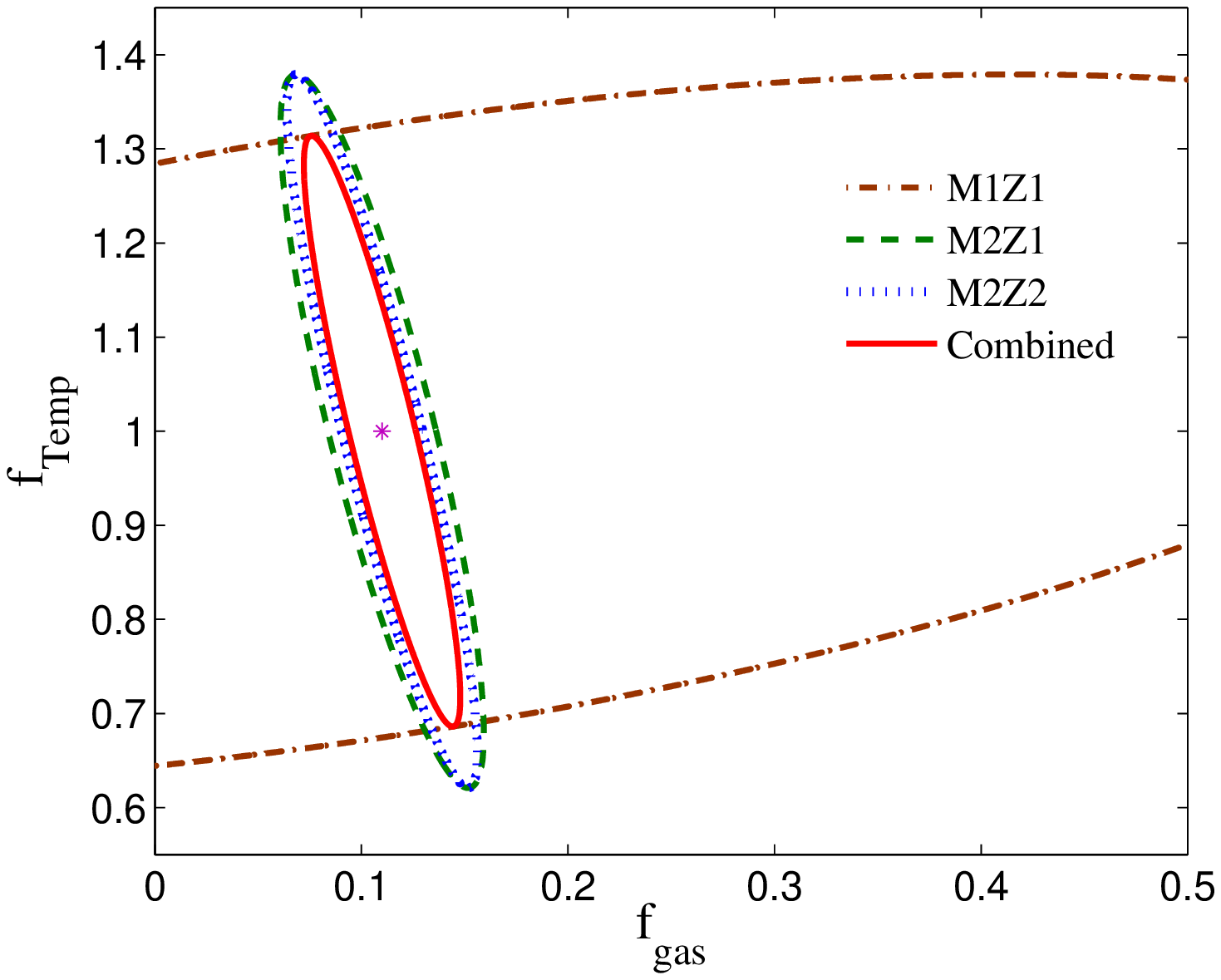}
 \includegraphics[width=8cm,angle=0.0 ]{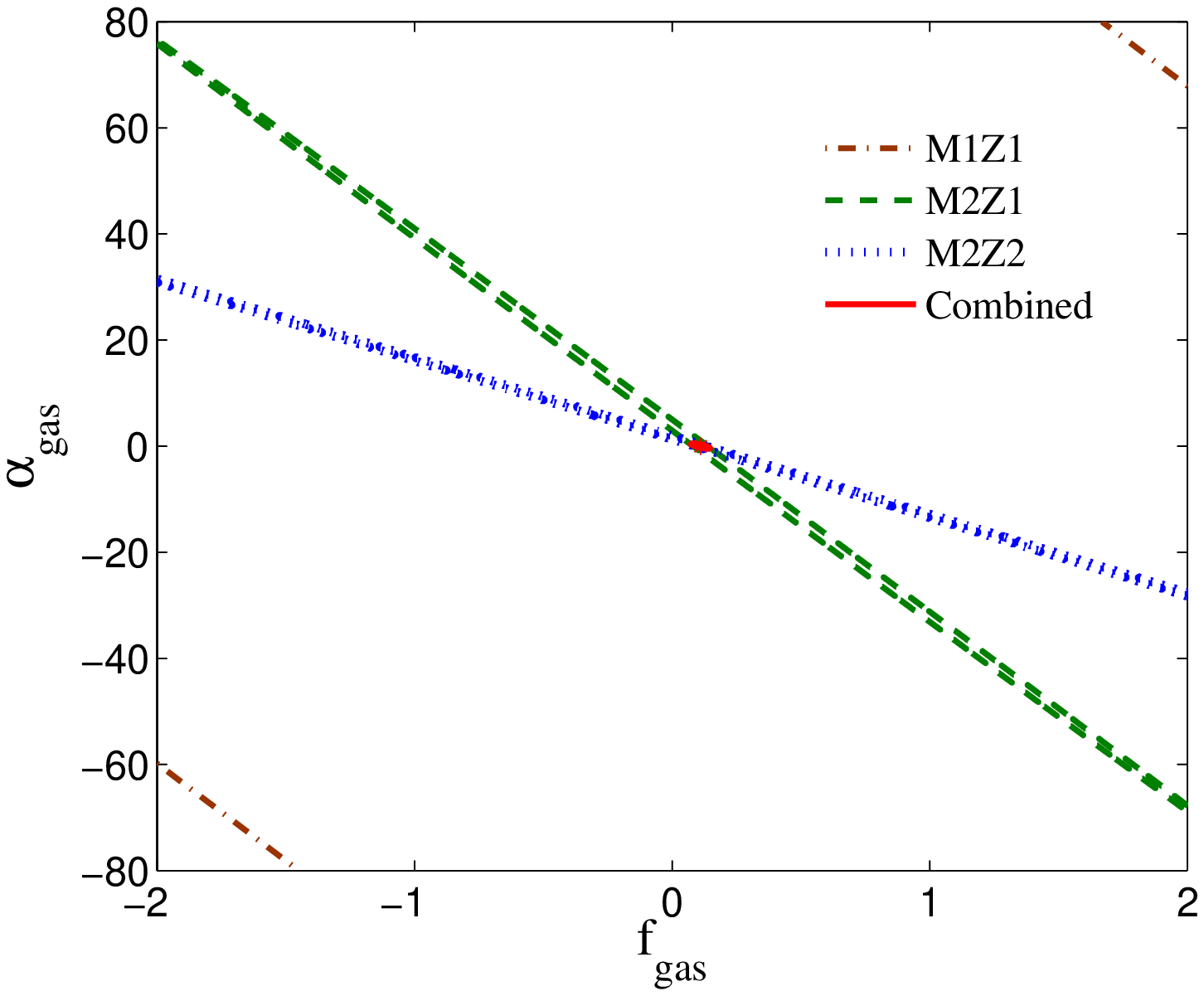}
 \includegraphics[width=8cm,angle=0.0 ]{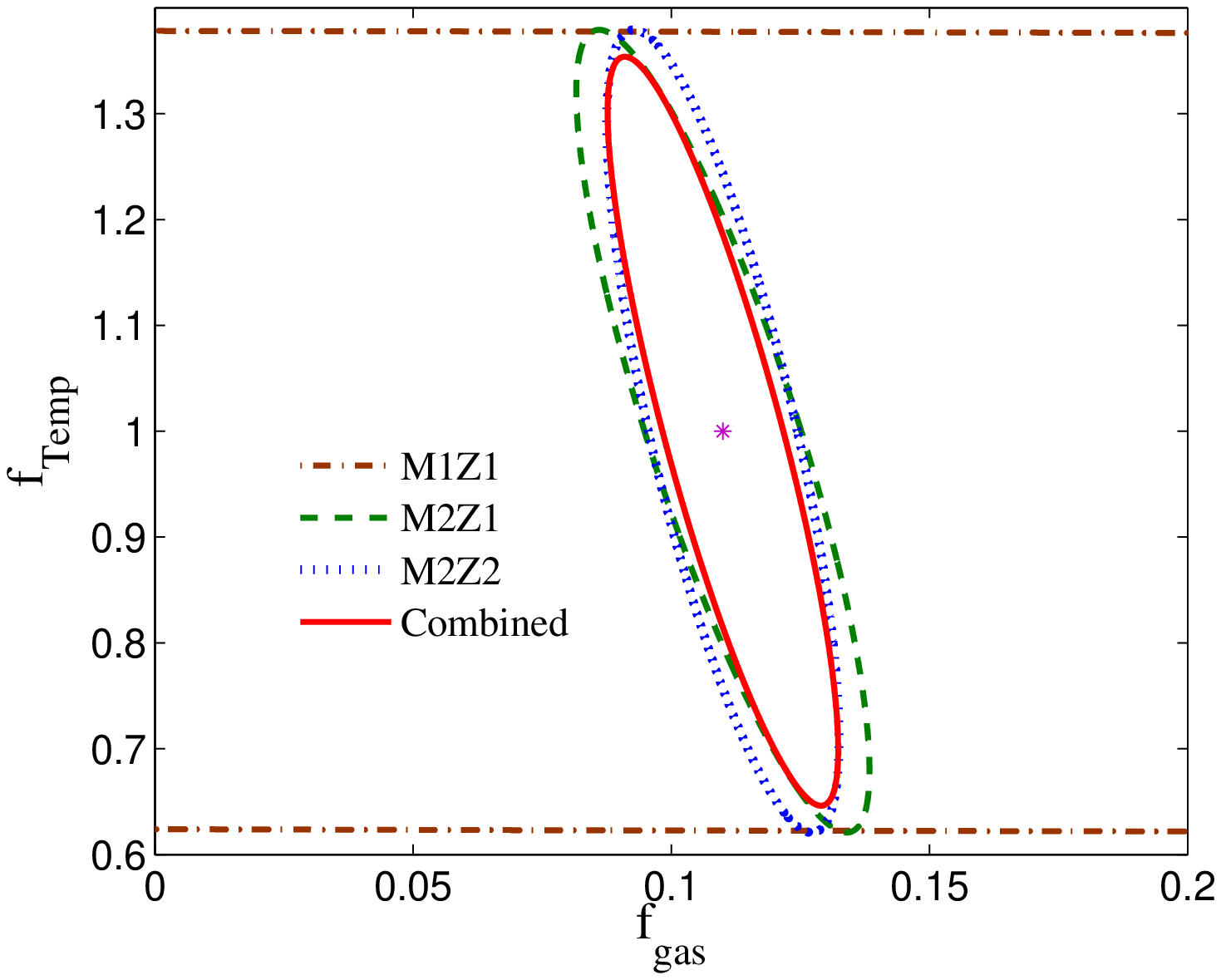}
 \caption {The 68\% CL for the gas physics parameters \fg, \ftemp and \alpg from the SPT-DES (SZ-galaxy cross power spectrum) and eROSITA-DES (X-ray-galaxy cross power spectrum)
surveys in the upper and lower panels respectively. All the plots are for \prb case and in the right panels, we have fixed \alpg. 
The brown (dot-dashed), green (dashed), blue (dotted) and red (solid) lines represent the ellipses for the mass and redshift 
bins M1Z1, M2Z1, M2Z2 and the combined ellipse respectively.}
 \label{fig-fisher}
\end{center}
\end{figure*}

Given a fiducial model, the Fisher matrix can be written as
\begin{equation}
 F_{ij} \,=\Sigma_{ll'} \,\frac{\partial C_\ell}{\partial p_i} (M_{\ell \ell{^\prime}})^{-1} \frac{\partial C_{\ell{^\prime}}}{\partial p_j} \delta_{ll'}
\end{equation}
where $M_{ll'}$ is the covariance matrix given by equation-(\ref{mll}). To calculate the uncertainty on the parameters we have considered 
following two prior cases:\\
\pra:  Priors on cosmological parameters only.\\
\prb:  Priors on cosmological parameters + 25\% prior on $f_{\rm Temp}$.

We have included only those galaxies for which the gas cooling time is more than the halo destruction time ensuring that the 
CGM temperature is close to the virial temperature of the halo. Therefore, we assume a small uncertainty in \ftemp in
\prb.


\begin{table}
\caption{Error on parameters for different scenarios (see Table-\ref{tab_fiducial}) for SPT-DES combination}
\centering 
\begin{tabular}{l l l l l }
\hline
 Parameter &  P1 &  P2 & P1 (fixed \alpg)& P2 (fixed \alpg)  \\
 \hline
 $ f_{\rm gas}$ & 0.049 & 0.037 & 0.042 & 0.025 \\[-0.5ex]
 $ f_{\rm Temp}$ & 0.369 & 0.207 & 0.369 & 0.207 \\[-0.5ex]
 $ \alpha_{\rm{gas}}$ & 0.519 & 0.519 & \hspace{3mm}- & \hspace{3mm}- \\[-0.5ex]
  \hline
 \end{tabular}
 \label{tab:yh}
\end{table}



\begin{table}
\caption{Error on parameters for different scenarios (see Table-\ref{tab_fiducial}) for eROSITA-DES combination}
\centering 
\begin{tabular}{l l l l l }
\hline
 Parameter &  P1 &  P2 & P1 (fixed \alpg)& P2 (fixed \alpg)  \\
 \hline
  $f_{\rm gas}$ & 0.20 & 0.025 & 0.036 & 0.015 \\[-0.5ex]
 $ f_{\rm Temp}$ & 2.65 & 0.25 & 0.649 & 0.233 \\[-0.5ex]
 $ \alpha_{\rm{gas}}$ & 1.219 & 0.318 & \hspace{3mm}- & \hspace{3mm}- \\[-0.5ex]
  \hline
 \end{tabular}
 \label{tab:xh}
\end{table}


\subsection{Results}
\label{sec-result}
In Table-(\ref{tab:yh}) and (\ref{tab:xh}) we show the forecasted uncertainty on the parameters for the SPT-DES and eROSITA-DES surveys
respectively. Here we have combined the Fisher from three different combinations of the mass and redshift bins M1Z1, M2Z1 and M2Z2
where M1=$[10^{11.8},10^{12.0}] h^{-1}M_{\odot}$, M2=$[10^{12.8},10^{13.0}] h^{-1}M_{\odot}$, Z1=$[0.4,0.5]$ and Z2=$[1.0,1.1]$.
We show the constraints on the astrophysical parameters only as the cosmological parameters are already well constrained by {\it{Planck}}.
There is a strong degeneracy between the gas physics parameters if we consider only one mass and redshift bin. However, when the
information from two or more bins are added together, we can break this degeneracy and obtain  strong constraints on these 
parameters. In Table-(\ref{tab:yh}) we show the constraints on the parameters from the SZ-galaxy cross-correlation signal for the SPT-DES survey.
Combining the Fisher matrix from M1Z1, M2Z1 and M2Z2 can constrain \fg to $\sim 44\%$
and \ftemp to $\sim 37\%$ around their fiducial values, even without any prior knowledge on astrophysical parameters. For \pra, the constraint on 
\alpg is $\Delta \alpha_{\rm{gas}} \sim 0.5$. Including additional $25\%$ prior on gas temperature does not improve
the constraint on \alpg whereas the constraint on \fg(\ftemp) improves considerably to $34\% (21\%)$. 

In the absence of any redshift evolution of the gas fraction, the constraint on \ftemp does not improve 
whereas the constraint on \fg improves to $\sim 38\%$ for \pra and to $\sim 23\%$ for \prb.

We use only above three mass-redshift bins to forecast the constraints on gas physics parameters as the addition of
more bins does not improve these constraints much. For example, for \pra, the addition of mass-redshift bin M2Z3, where  
Z3=$[0.8,0.9]$, improves the constraint on \fg and \alpg from 0.049 and 0.5 to 0.047 and 0.46 respectively whereas the change
in the constraint on \ftemp is $<$1\%. Also the change in these constraints in other prior cases is negligible. Further addition
of mass-redshift bins in the Fisher matrix analysis does not improve these constraints. Therefore, we use only the bins M1Z1,
M2Z1 and M2Z2 for the purpose of our work.

In Table-(\ref{tab:xh}) we show the constraints on the parameters from the X-ray-galaxy cross-correlation signal for the
eROSITA-DES survey. Now for \pra and in the presence of unknown redshift evolution of the  gas fraction, astrophysical parameters
are poorly constrained by this survey. This is mainly due to the large noise contamination from the X-ray background. The addition of a  $25\%$
prior on \ftemp significantly improves the constraints on the parameters. The uncertainty on \fg reduces
to $23\%$ and \alpg can be constrained to $\Delta \alpha_{\rm{gas}} \sim 0.3$.

In the absence of any redshift evolution of the gas fraction, \fg(\ftemp) can be constrained to $\sim 33\%
(65\%)$ for \pra and the constraint becomes $\sim 14\% (23\%)$ for \prb.

In Figure-(\ref{fig-fisher}) we show the constraints from the SPT-DES (SZ-galaxy cross power spectrum) and eROSITA-DES (X-ray-galaxy cross power spectrum)
surveys in the upper and lower panels respectively. All the plots are for \prb and  in the right panels, we have fixed \alpg.
In the upper left panel we show the 68\% confidence limit (CL) ellipse for \fg and \alpg for the SZ-galaxy cross power spectrum. The individual
ellipses for M1Z1, M2Z1 and M2Z2 are large and  there is a large uncertainty on these parameters. But combining them together results in
$\Delta f_{\rm {gas}}\approx 0.037$ i.e. we can constrain \fg to $\sim 34\%$. This is because the degeneracy of \fg
with \alpg is broken when we add information from the galaxies in similar mass bins but in different redshift bins.
The X-ray-galaxy cross power spectrum also has similar contours for \fg v/s \alpg (lower left panel of Figure-(\ref{fig-fisher})).
In this case, the Fisher matrix analysis gives a constraint $\Delta f_{\rm {gas}}\approx 0.025$ on gas fraction, i.e. $23\%$ of its fiducial value.
In both the cases, M1Z1 bin has relatively large uncertainty and the final uncertainty ellipse is essentially determined by the other two
bins. The amount of gas present in the M1Z1 bin is roughly an order of magnitude smaller than that of the other bins which results
in a smaller signal and large uncertainty on the parameters.

In the upper right panel of Figure-(\ref{fig-fisher}) we show the constraints on the SPT-DES survey for \prb case and the redshift-independent gas fraction. 
Here adding the information from different bins does not improve the constraints on the parameters
as we already have a strong prior on \ftemp ($\sim 25\%$ of its fiducial value). However, even with a strong prior on the
CGM temperature, there is a large uncertainty in the gas fraction for M1Z1 bin which is again due the small signal in this bin.
In the lower right panel on Figure-(\ref{fig-fisher}) we show the 68\% CL ellipses for \fg-\ftemp from eROSITA-DES survey.
This survey can constrain the gas fraction to $\Delta f_{\rm{gas}}\approx 0.015$ in case of a redshift-independent \fg.

\begin{figure}
\begin{center}
 \includegraphics[width=9.0cm,angle=0.0 ]{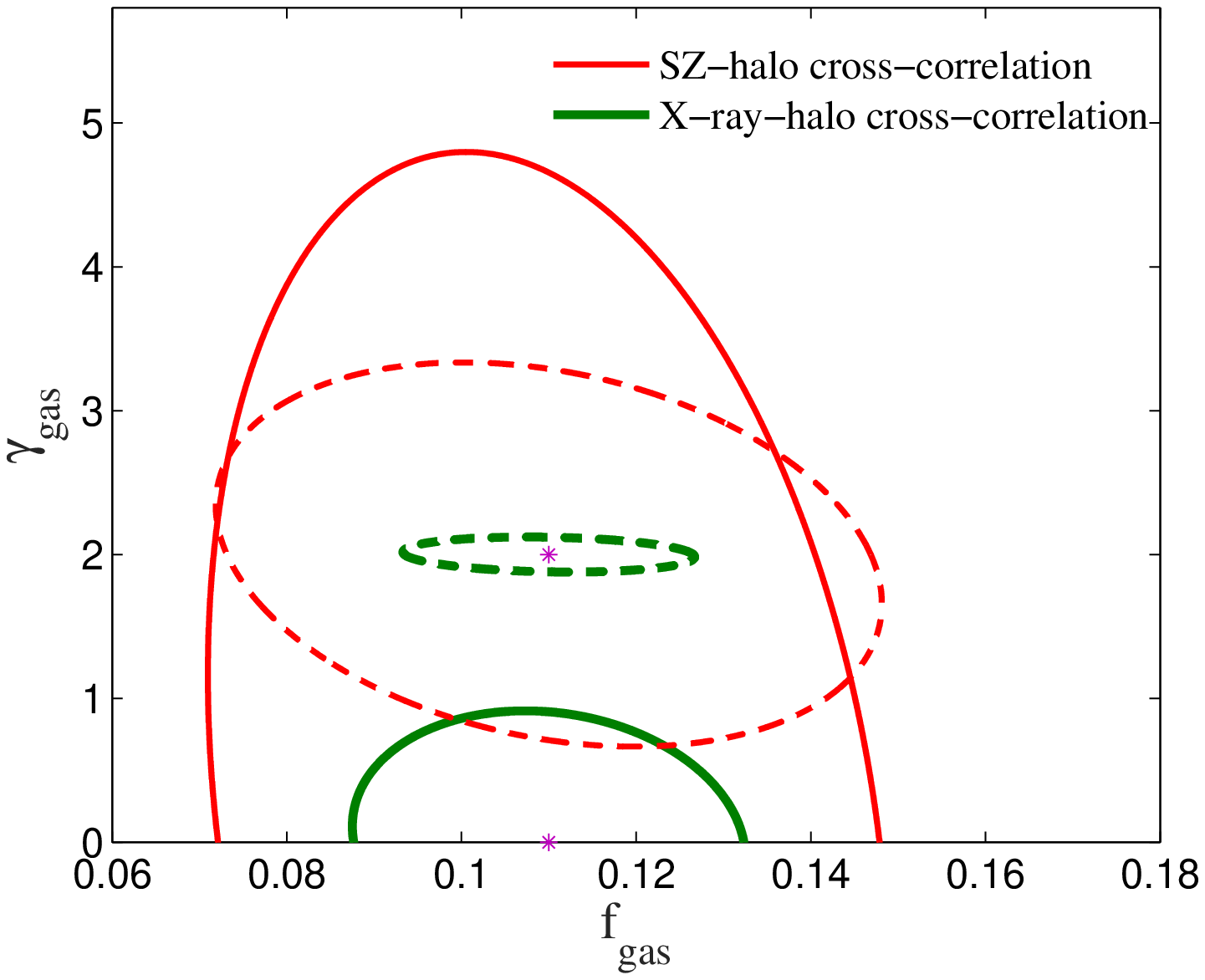}
 \caption {The 68\% CL contours for \fg and \gmag from the SPT-DES (SZ-galaxy cross power spectrum) and eROSITA-DES 
 (X-ray-galaxy cross power spectrum) surveys. The solid thin red (thick green) ellipse is for the fiducial value \gmag=0 for the SZ (X-ray)-galaxy cross power spectrum
 and the dashed thin red (thick green) ellipse is for the fiducial value \gmag=2 for SZ (X-ray)-galaxy cross power spectrum.
 All the plots are for \prb case with fixed \alpg.}
 \label{fig-gamma}
\end{center}
\end{figure}

Note that, in the calculation of the constraints on gas physics parameters, we have used strong priors on the cosmological 
parameters (from \cite{planck15}). However, if we remove the prior on the cosmological parameters, it increases the uncertainty
in the determination of the gas physics parameters. For example, if we remove the prior on $\sigma_8$, the uncertainty on the 
gas physics parameters becomes larger than 100\% for SZ-galaxy as well as X-ray-galaxy cross-correlation in \pra case. But the 
inclusion temperature prior (\prb case) improves the constraint on \fg ($\Delta$\fg = 0.041 (0.026) for SZ-galaxy 
(X-ray-galaxy) cross-correlation) and \alpg ($\Delta$\alpg = 0.52 (0.38) for SZ-galaxy (X-ray-galaxy) cross-correlation)
significantly even in the absence of any prior
on $\sigma_8$. These constraints are similar to the constraints on these parameters in the presence of {\it{Planck}} prior
on $\sigma_8$. Therefore, the prior knowledge of one of the gas physics parameter helps in breaking the degeneracy between
the cosmological and gas physics parameters.

\section{Probing the energetics of the CGM}
\label{sec-feedback}
Currently, there is a large uncertainty in the knowledge of the distribution of the CGM. Simulations suggest that
the extent and distribution of the CGM  also depends on the feedback processes taking place in the central part of the galaxy
\citep{brun15, joshua15}. Without any feedback, the temperature at outer radii falls rapidly, whereas the winds and AGN feedback tend to make the profile flatter. The effect of the feedback on the density profile is however weaker than on the temperature profiles. In other words, the pressure profile is likely to be flatter at the outer radii in the presence of the feedback processes than
in the cases without any feedback.

In this section, we investigate whether the SZ/X-ray-galaxy cross power spectrum can distinguish between different density profiles of the CGM and 
hence the processes giving rise to these profiles. In order to constrain the density profile of the CGM, assuming the gas to be at the virial temperature, we now 
include \gmag in the Fisher matrix analysis, where \gmag is defined by
\begin{equation}
 \rho_{gas}=\rho_0 [1+(r/R_s)^{\gamma_{\rm{gas}}}]^{-1}
 \label{eqn-density}
\end{equation}
where $\rho_0$ is the normalization such that the mass within the virial radius of the galaxy remains the same and the fiducial
value of \gmag is 0 for the flat density profile. For the SPT-DES survey,
the uncertainty on the \gmag is large and it can at best be constrained to \gmag$<3.16$ in the absence of any
redshift evolution of \fg and \prb case. Even when we include the redshift evolution of gas fraction and remove the prior
on gas temperature, the uncertainty on \gmag degrades only slightly to \gmag$<3.35$. This shows that the 
SZ-galaxy cross power spectrum is less sensitive to the density profile of the CGM as compared to other astrophysical parameters within the resolution of 
the SPT ($l\sim 10^4$). Varying the density profile affects the SZ-galaxy cross power spectrum only at large $l$-values and hence
this situation can only be improved by a higher resolution CMB survey in the future.

On the other hand, the X-ray emission which is proportional to the square of the density of the CGM, is
much more sensitive to its density profile even at small $l$-values. Also the resolution of the eROSITA ($l \sim 30000$) is much
better than the resolution of the SPT. As a result, \gmag can be constrained to \gmag$<0.6$ by the X-ray-galaxy cross power 
spectrum. This constraint remains almost invariant even if we fix \alpg in the \prb case.

In the case of a steeper density profile i.e. a larger value of \gmag, both the SZ-galaxy and X-ray-galaxy cross power spectrum
increase, specially at large $l$-values (see Figure-(\ref{fig-prf}) and (\ref{fig-xprf})). This results in an increased signal-to-noise ratio, as well as an improvement in the
constraints on the parameters. For example, if we take \gmag=2, the constraint on \gmag improves to 
$\Delta \gamma_{\rm{gas}}<1$ from the SZ-galaxy cross power spectrum whereas the slope of the density profile can be constrained 
with an accuracy better than 5\% (i.e. $\Delta \gamma_{\rm{gas}}<0.1$) from the X-ray-galaxy cross power spectrum.

In Figure-(\ref{fig-gamma}), we show the 68\% CL ellipses for \fg-\gmag from the SPT-DES and 
the eROSITA-DES surveys. Again, due to the more sensitive dependence of the X-ray emission on the CGM density profile as compared to the SZ-effect, the uncertainty
ellipses of the X-ray-galaxy cross-correlation are smaller and hence can constrain \gmag better than the SZ-galaxy cross-correlation.

This result has implications of being able to constrain the distribution and evolution of the CGM.
Recent simulations \citep{brun15, joshua15} have shown that the feedback processes (from star formation and AGNs) can affect 
the density and temperature profiles of the CGM. These simulations match some of the observed properties of the CGM and galaxies.
However, they are quite sensitive to the feedback mechanism used and give a variety of the CGM density and temperature profiles 
depending on the feedback process. The density profiles of the CGM in these simulations can be reasonably fit 
by $1.2 \le \gamma_{\rm{gas}} \le 2.5$, within the virial radius of the galaxy for various feedback processes. For example, the No AGN, fixed-v hot
winds and fully enriched winds models of \cite{joshua15} can be fit with \gmag=2, 1.6 and 2.2 respectively (excluding
the central part). Also the pressure profile for the massive galaxies from \cite{brun15} (see the first panel of Figure-(3) of \cite{brun15}) can be fit with
a \gmag$ \sim 1.25 $, assuming the gas to be at the virial temperature. Therefore the X-ray-galaxy and SZ-galaxy cross power spectrum have the potential of
discriminating between the evolutionary processes for the CGM, at greater than $3\sigma$,
if $\Delta \gamma_{\rm{gas}}$ can be constrained within $\sim 0.5$.

\section{Conclusions}
\label{sec-conc}
We have studied the cross-correlation power spectra for the SZ-galaxy distribution, X-ray-galaxy distribution and X-ray-SZ effect for the
hot gas in the galactic haloes. Our main conclusions are as follows: 

\begin{enumerate}
  \item We predict that the SZ-galaxy cross power spectrum is significant at small scales ($l\gtrsim3000$) and can be
detected at SNR $\sim9$ by combining the SPT and DES surveys for the massive galaxies at intermediate redshifts. 
The shape of the SZ-galaxy cross-power spectrum is sensitive to the underlying distribution of the gas at large 
$l$-values ($l\approx 10^4$).

  \item For the X-ray-galaxy cross power spectrum, we have considered the combination of the eROSITA-DES and eROSITA-LSST surveys and these 
surveys can detect the signal at SNR$\sim6$ and $12$ respectively for the high mass and intermediate redshift galaxies. For the 
flat density profile, the shape of the X-ray-galaxy cross power spectrum is similar to the shape of the SZ-galaxy cross power spectrum. However,
the X-ray emission ($\propto n^2_e$) is more sensitive to the density profile than the SZ-effect ($\propto n_e$). As a result, the 
X-ray-galaxy cross power spectrum vaires significantly with the steepening of the density profile even at $l\approx3000$.

 \item The possibility of detecting the X-ray-SZ cross power spectrum from the CGM is low (SNR $< 1$) for 
the SPT-eROSITA combination. This is due to the combined effect of the high noise in X-rays and low resolution of the SPT survey.

  \item Finally, we do a Fisher matrix analysis for these surveys to forecast the constraints that can be derived on the 
  amount of gas in the CGM. After 
  marginalizing over the cosmological parameters with $Planck$ priors and combining the Fisher matrix analysis for three different mass and redshift
bin combinations, the SPT-DES survey can constrain \fg to $\sim 34$\% in the presence, and to $\sim 23$\% in the absence,
of any possible redshift evolution of the gas fraction. For the same set of mass and redshift bins, the  eROSITA-DES survey can constrain
\fg to $\sim 23$\% and $\sim 14$\% in the presence and absence of redshift evolution of gas fraction respectively.
Note that we neglect the correlation between the galaxies and the AGNs in the calculation of the uncertainties in the 
X-ray-galaxy cross-power spectrum.

  \item Including the slope of the density profile \gmag (defined in Eq.\ref{eqn-density}) in the 
Fisher matrix analysis, degrades the constraints on other astrophysical parameters whereas \gmag itself can be constrained
to \gmag$ < 0.6$ ($< 3.4$) by the X-ray (SZ)-galaxy cross power spectrum for the flat density profile. These constrains
are sensitive to the fiducial value of \gmag and improve for a steeper density profile of the CGM. For \gmag=2,
it can be constrained to $\Delta \gamma_{\rm{gas}}<0.1$ ($<1.0$) by the X-ray (SZ)-galaxy cross power spectrum power spectrum.

 \item In all our calculations, we have assumed \fg=0.11. If instead, we take a low value of the gas fraction in the CGM, e.g. \fg$\approx0.05$,
the SZ cross-power spectrum and its detectability goes down roughly by a factor 2 as the SZ signal is proportional to the amount of
gas. So with \fg$\approx0.05$, the peak SNR $\sim 4$-$5$ for the SPT-DES survey.
However, since the X-ray signal is proportional to $n^2_e$, the X-ray cross power spectrum and its detectability goes down
by a factor of 4. As a result, the SNR goes below $2$ for the eROSITA-DES survey whereas this signal can still be detected 
at SNR $\sim 3$ with the eROSITA-LSST combination. 
  
\end{enumerate}

Presently, the amount of CGM in galactic haloes, its distribution, energetics and other properties are not well determined. Therefore, the 
detection and study of the SZ-galaxy and X-ray-galaxy cross power spectrum can provide powerful constraints on the nature of the CGM and open up
the possibility of differentiating between various feedback models which affect the evolution of the CGM.\vspace{10mm}\\
{\bf{ACKNOWLEDGEMENTS}}\\
We thank the anonymous referee for valuable suggestions and comments.
PS thanks Biswajit Paul, Kartick C. Sarkar, Nazma Islam and Nafisa Aftab for helpful discussions.
We would like to thank Joshua Suresh for useful discussions.

\footnotesize{

}


\begin{thebibliography}{}

\bibitem[\protect\citeauthoryear{Anderson 
\& Bregman}{2010}]{anderson10} Anderson, M.~E., Bregman, J.~N., 2010, ApJ, 714, 320 

\bibitem[\protect\citeauthoryear{Anderson \& Bregman}{2011}]{anderson11}
Anderson, M. E., Bregman, J. N. 2011, ApJ, 737, 22

\bibitem[\protect\citeauthoryear{Anderson \etal}{2013}]{anderson13}
Anderson, M. E., Bregman, J. N., Dai, X. 2013, ApJ, 762, 106

\bibitem[\protect\citeauthoryear{Bogd\'an \etal}{2013a}]{bogdan13a}
Bogd\'an, \'A, Forman, W. R., Vogelsberger, M. \etal 2013, ApJ, 772, 97

\bibitem[\protect\citeauthoryear{Bogd\'an \etal}{2013b}]{bogdan13b}
Bogd\'an, \'A, Forman, W. R., Kraft, R. P., Jones, C. 2013, ApJ, 772, 98

\bibitem[\protect\citeauthoryear{Cheng \etal}{2003}]{cheng03}
Cheng, L.-M., Wu, X.-P., Cooray, A. 2003, A\&A, 413, 65c

\bibitem[\protect\citeauthoryear{Dai \etal}{2012}]{dai12}
Dai, X., Anderson, M. E., Bregman, J. N., Miller, J. M. 2012, ApJ, 755, 107

\bibitem[\protect\citeauthoryear{Diego \etal}{2003}]{diego03}
Diego, J. M., Silk, J., Sliwa, W. 2003, MNRAS, 346, 940

\bibitem[\protect\citeauthoryear{Duffy \etal}{2008}]{duffy08}
Duffy, A. R., Battye, R. A., Davies, R. D., Moss, A., Wilkinson, P. N. 2008, MNRAS, 383, 150

\bibitem[\protect\citeauthoryear{Dutton \etal}{2010}]{dutton10}
Dutton, A. A., Conroy, C., vanden Bosch, F. C., Prada, F., More, S. 2010, MNRAS, 407, 2D

\bibitem[\protect\citeauthoryear{Fang \etal}{2012}]{fang12}
Fang, W., Kadota, K., Takada M. 2012, PhRvD, 85b3007F

\bibitem[\protect\citeauthoryear{Gatto \etal}{2013}]{gatto13} Gatto, A., Fratfernali, 
F., Read, J.~I., et al.\ 2013, MNRAS, 433, 2749 

\bibitem[\protect\citeauthoryear{Grcevich \& Putman}{2009}]{putman09}
Grcevich, J., Putman, M. E. 2009, ApJ, 696, 385

\bibitem[\protect\citeauthoryear{Hajian \etal}{2013}]{hajian13}
Hajian, A., Battaglia, N., Spergel, D. N., Bond, J. R., Pfrommer, C., Sievers, J. L. 2013, JCAP, 11, 064

\bibitem[\protect\citeauthoryear{Hern\'{a}ndez-Monteagudo \etal}{2004}]{hern04}
Hern\'{a}ndez-Monteagudo, C., Genova-Santos, R., Atrio-Barandela, F. 2004, ApJ, 613, L89

\bibitem[\protect\citeauthoryear{Hern\'{a}ndez-Monteagudo \etal}{2006}]{hern06}
Hern\'{a}ndez-Monteagudo, C., Mac\'{\i}as-P\'{e}rez, J. F., Tristram, M., D\'{e}sert, F.-X. 2006, A\&A, 449, 41

\bibitem[\protect\citeauthoryear{Hinshaw \etal}{2007}]{hinshaw07}
Hinshaw, G., Nolta, M. R., Bennett, C. L., et al. 2007, ApJS, 170, 288

\bibitem[\protect\citeauthoryear{Hurier \etal}{2014}]{hurier14}
Hurier, G., Aghanim, N., Douspis, M. 2014, A\&A, 568, A57

\bibitem[\protect\citeauthoryear{Hurier \etal}{2015}]{hurier15}
Hurier, G., Douspis, M., Aghanim, N., Pointecouteau, E., Diego, J. M., Macias-Perez, J. F. 2015, A\&A, 576, A90

\bibitem[\protect\citeauthoryear{Komatsu \& Kitayama}{1999}]{komatsu99}
Komatsu, E., Kitayama, T. 1999, ApJ, 526L, 1K

\bibitem[\protect\citeauthoryear{Leauthaud \etal}{2010}]{leau10}
Leauthaud, A., Tinker, J., Bundy, K., et al. 2012, ApJ, 744,159

\bibitem[\protect\citeauthoryear{Le Brun \etal}{2015}]{brun15}
Le Brun, M. M. C., McCarthy, I. G., Melin, J. B. 2015, MNRAS, 451, 3868

\bibitem[\protect\citeauthoryear{Ma \etal}{2014}]{ma14}
Ma, Y.-Z., Van Waerbeke, L., Hinshaw, G., Hojjati, A., Scott, D., Zuntz J., 2015, J. Cosmology Astropart. Phys., 9, 46

\bibitem[\protect\citeauthoryear{Merloni \etal}{2012}]{merloni12}
Merloni, A., Predehl, P., Becker, W., et al. 2012, arXiv, 1209.3114

\bibitem[\protect\citeauthoryear{Miller \& Bregman}{2015}]{miller15} Miller M.~J., Bregman J.~N., 2015, ApJ, 800, 14 

\bibitem[\protect\citeauthoryear{Mo \etal}{1998}]{mo98}
Mo, H. J., Mao, S., White, S. D. M. 1998, MNRAS, 295, 319

\bibitem[\protect\citeauthoryear{Moster \etal}{2010}]{moster10}
Moster, B. P., Maccio, A. V., Somerville, R. S., Johansson, P. H., Naab, T. 2010, MNRAS, 403, 1009M

\bibitem[\protect\citeauthoryear{Oguri \& Takada}{2011}]{oguri11}
Oguri, M., Takada, M. 2011, Phys. Rev D 83, 023008 

\bibitem[\protect\citeauthoryear{Planck Collaboration XI}{2013}]{planck13}
Planck Collaboration 2013, A\&A, 557, 52

\bibitem[\protect\citeauthoryear{Planck results XIII}{2015}]{planck15}
Planck results XIII 2015, arXiv, 1502.01589

\bibitem[\protect\citeauthoryear{Planck results XXIV}{2015}]{planck15a}
Planck results XXIV 2015, arXiv, 1502.01597

\bibitem[\protect\citeauthoryear{Putman \etal}{2012}]{putman12}
Putman, M. E., Peek, J. E. G., Joung, M. R. 2012, ARA\&A, 50, 491

\bibitem[\protect\citeauthoryear{Rines \etal}{2015}]{rines15}
Rines, K. J., Geller, M. J., Diaferio, A., Hwang, H. S. 2015, arXiv, 1507.08289

\bibitem[\protect\citeauthoryear{Sheth \& Tormen}{1999}]{sheth99}
Sheth, R. K, Mo, H. J., Tormen, G. 1999, MNRAS, 308, 119

\bibitem[\protect\citeauthoryear{Sheth \& Tormen}{2001}]{sheth01}
Sheth, R. K, Mo, H. J., Tormen, G. 2001, MNRAS, 323, 1S

\bibitem[\protect\citeauthoryear{Singh \etal}{2015}]{singh15}
Singh, P., Nath, B. B., Majumdar, S., Silk, J. 2015, MNRAS, 448, 2384S

\bibitem[\protect\citeauthoryear{Sunyaev \& Zel'dovich }{1969}]{sz69}
Sunyaev, R. A., Zel'dovich, Ya. B. 1969, Nature, 223, 721

\bibitem[\protect\citeauthoryear{Suresh \etal}{2015}]{joshua15}
Suresh, J., Bird, S., Vogelsberger, M. \etal 2015, MNRAS, 448, 895

\bibitem[\protect\citeauthoryear{Sutherland \& Dopita}{1993}]{dopita93}
Sutherland, R. S., Dopita, M. A. 1993, ApJ, 88, 253S

\bibitem[\protect\citeauthoryear{DES collaboration}{2015}]{des15}
The Dark Energy Survey Collaboration, Abbott, T., Abdalla, F. B., et al. 2015, arXiv, 1507.05552

\bibitem[\protect\citeauthoryear{Van Waerbeke \etal}{2014}]{van14}
Van Waerbeke, L., Hinshaw, G., Murray, N. 2014, PRD, 89, 023508 
   
\bibitem[\protect\citeauthoryear{Werk et al.}{2014}]{werk14} 
Werk, J.~K., et al., 2014, ApJ, 792, 8 


\end{thebibliography}
\end{document}